\documentclass[journal, onecolumn, 11pt]{IEEEtran}

\usepackage{verbatim}
\usepackage{graphicx}
\usepackage[cmex10]{amsmath}
\usepackage{amssymb}
\usepackage[caption=false,font=footnotesize]{subfig}
\usepackage{url}
\usepackage{color}

\def\modifiedflag{1}
\ifx\modifiedflag\undefined

\else

\fi

\newtheorem{thm}{Theorem}
\newtheorem{defn}{Definition}
\newtheorem{lem}{Lemma}
\newtheorem{indef}{Informal Definition}

\begin{document}

\title{\textbf{Myopic Coding in Multiterminal Networks}}
\author{Lawrence Ong, {\em Student Member, IEEE} and Mehul Motani, {\em Member, IEEE} \\
\thanks{A portion of the results in this paper has been presented at the 39th Conference on Information Sciences and Systems, John Hopkins University, Baltimore, MD, 16-18 March, 2005, and the IEEE International Symposium on Information Theory, Adelaide Convention Centre, Adelaide, Australia, 4-9 September, 2005.}
Electrical and Computer Engineering Department,\\
National University of Singapore, Singapore 119260.\\
Email: lawrence.ong@cantab.net, motani@nus.edu.sg} 


\maketitle

\begin{abstract}
This paper investigates the interplay between cooperation and achievable rates in multi-terminal networks. Cooperation refers to the process of nodes working together to relay data toward the destination. There is an inherent tradeoff between achievable information transmission rates and the level of cooperation, which is determined by how many nodes are involved and how the nodes encode/decode the data. We illustrate this trade-off by studying information-theoretic decode-forward based coding strategies for data transmission in multi-terminal networks. Decode-forward strategies are usually discussed in the context of {\em omniscient coding}, in which all nodes in the network fully cooperate with each other, both in encoding and decoding. In this paper, we investigate {\em myopic coding}, in which each node cooperates with only a few neighboring nodes. We show that achievable rates of myopic decode-forward can be as large as that of omniscient decode-forward in the low SNR regime. We also show that when each node has only a few cooperating neighbors, adding one node into the cooperation increases the transmission rate significantly.   Furthermore, we show that myopic decode-forward can achieve non-zero rates as the network size grows without bound.
\end{abstract}

{\keywords Achievable rates, decode-forward, multiple-relay channel, multi-terminal network, myopic coding.}

\section{Introduction}
\subsection{Wireless Networks}
Wireless networks have been receiving much attention recently by both researchers and industry.  The main advantage of wireless technology to users is the seamless access to the network whenever and wherever they are; to service providers, easier deployment, as no cable laying is required.  Examples of wireless networks include cellular mobile networks, Wi-Fi networks, and sensor networks.  A large amount of research has been carried out recently on various aspects of wireless networks, including power saving \cite{yukrishnamachari04,younisfahmy04}, routing \cite{shakkottai04,fanggao04,zhaoliu03}, transport capacity \cite{gopalaelgamal04,guptakumar03}, and connectivity \cite{shakkottaisrikant03}. In this paper, we focus on transmission rates in multi-terminal wireless networks.

Analyzing transmission rates in multi-terminal networks is not easy. Consider the \emph{single-relay channel}~\cite{covergamal79,meulen71}, a channel consisting of one source, one relay, and one destination. Even for this simple three-terminal network, the capacity is not known except for a few special cases, e.g., the degraded relay channel~\cite{covergamal79}. This hints at the difficulty of analyzing multi-terminal networks. We attempt to investigate an excerpt of the multi-terminal network by looking at data transmission from a single source to a single destination, from multiple sources to a single destination, and from a single source to multiple destinations, with the help of relay(s). Appropriate models for these types of networks are the \emph{multiple-relay channel}~\cite{aref80,xiekumar03} (an extension of the single-relay channel), the \emph{multiple-access relay channel} \cite{sankara04,sankara04b}, and the \emph{broadcast relay channel} \cite{liang04} respectively.
The reason for using relays, which have no data of their own to send, in the network is as follows.
Direct transmission from the source to a far-situated destination may require high transmission power (due to the path loss of electromagnetic wave propagation).
Since wireless networks operate over a shared medium,  this can create direct interference to other users.
Transmitting data via intermediate relays, using multiple-hop routing or cooperative relaying, can help to decrease the transmit power and reduce multi-user interference.

\subsection{Point-to-Point Coding}
A common approach to data transmission is to abstract the wireless network into a communication graph, with an edge connecting two nodes if they can communicate. Data communication happens by identifying a route, which is a sequence of nodes that connect the source to the destination. Each node sends data to the next node in the route and decodes data from the previous node in the route. Transmissions of other nodes are treated as noise. We call this coding strategy \emph{point-to-point coding} in a multi-terminal network. This way of transmitting data from the source to the destination is commonly called multi-hop routing in the communications and networking literature. The terms coding and coding strategy are used interchangeably in this paper.

\subsection{Omniscient Coding}
Point-to-point coding ignores the inherent broadcast nature of the wireless channel, i.e., that a node can hear transmissions meant for other nodes, and thus it can act as a relay for them.  Clearly, the best thing to do is for all nodes to cooperate, helping the source to send its data to the destination.  This requires every node to be aware of the presence of other nodes and to have knowledge of the processing they do. We refer to coding strategies that utilize the global view and complete cooperation as {\em omniscient coding}. In the literature, omniscient coding strategies were investigated for multi-terminal networks, e.g., the multiple-access relay channel, the broadcast relay channel \cite{kramer00,kramergastpar04b}, and the multiple-relay channel \cite{guptakumar03,xiekumar03,kramergastpar04}. While the rates achievable by omniscient coding strategies are higher than those achievable by point-to-point coding strategies in these channels, there are a number of practical difficulties in implementing complete cooperation, e.g., (i) designing codes based on omniscient coding is more difficult as it involves the optimization of the whole network, (ii) the failure of one node affects the decoding of all other nodes, and (iii) all nodes need to be synchronized (for some coding strategies).

\subsection{Myopic Coding}
In view of these practical issues, we investigate \emph{myopic coding}, coding strategies with constrained communications, e.g., node have a local view of the network, and limited cooperation. Myopic coding positions itself between point-to-point coding and omniscient coding. In myopic coding, communications of the nodes are constrained in such a way that a node communicates with more than two nodes (as opposed to point-to-point coding) but not with all the nodes (as opposed to omniscient coding) in the network. Myopic coding incorporates local cooperation. It allows cooperation among neighboring nodes to increase the transmission rate compared to point-to-point coding. On the other hand, it partially solves the practical difficulties encountered in omniscient coding.
In this paper, we illustrate myopic coding by using decode-forward based coding strategies.

We derive achievable rates of myopic coding strategies for the multiple-relay channel, the multiple-access relay channel, and the broadcast channel. We compare the performance of myopic coding to that of omniscient coding in these channels and show the trade-off between achievable rates and complexity.

\subsection{Contributions}
The primary aim of this work is to understand how to communicate
data from sources to destinations through a network of wireless relays.
This work is a step in the direction of designing efficient
protocols and algorithms for wireless networks. We ask the following questions which we will partially answer in the rest of this paper:
\begin{itemize}
\item What rate regions are achievable in multi-terminal channels (such as the multiple-relay channel, multiple-access relay channel, and the broadcast relay channel) in which every node has only a localized or myopic view of the network?
\item What is the value of cooperation?  In other words, what is the impact on the performance, in terms of transmission rates, when communications among the nodes are constrained compared to the case when they are unconstrained?
\end{itemize}

Answering these questions leads to the main contributions of this paper, which are:
\begin{itemize}
\item We construct random codes for \emph{myopic decode-forward}, i.e., decode-forward coding strategies~\cite{xiekumar03} with myopic outlook, for the discrete memoryless multiple-relay channel and derive achievable rates of the strategies.
\item We compute achievable rates of myopic decode-forward and omniscient decode-forward for the Gaussian multiple-relay channel.
\item Comparing the myopic version and the omniscient version of decode-forward, we show that including a few nodes into the cooperation increases the transmission rate significantly, often making it close to that under full cooperation.  In other words, sometimes more cooperation yields diminishing returns.
\item We show that in the multiple-relay channel, myopic decode-forward can achieve non-zero rates as the network size grows to infinity.
\item  We derive achievable rate regions of myopic decode-forward for the multiple-access relay channel and the broadcast relay channel.  On Gaussian channels, we show that under certain conditions, the performance of myopic coding can be close to that of omniscient coding.
\end{itemize}

\subsection{Paper Outline}
The rest of the paper is organized as follows.
In Section~\ref{sec:myopic-coding}, we define myopic coding and give examples of two myopic coding strategies. We present the advantages of myopic coding compared to omniscient coding. In Section~\ref{sec:myopic-mrc}, we investigate myopic coding in the multiple-relay channel. We first define the channel model and then derive achievable rates of two-hop myopic decode-forward. We then compare achievable rates of one-hop myopic decode-forward, two-hop myopic decode-forward, and omniscient decode-forward for the multiple-relay channel. We show that, in the five-node and the six-node Gaussian multiple-relay channels, when the nodes transmit at low signal-to-noise ratio (SNR), achievable rates of the two-hop coding are close to those of the omniscient coding. In Section~\ref{sec:k_hop}, we extend the analysis to the general $k$-hop myopic decode-forward for the $T$-node multiple-relay channel, where $k$ can be any positive integer from 1 to $T-1$ and $T$ is the number of nodes (including the source, the relays, and the destination) in the channel. In Section~\ref{sec:t_node}, we investigate myopic coding in a large network, meaning that the number of nodes grows to infinity. We show that even with a restricted view, in which a node treats the transmissions of the nodes beyond its view as noise, achievable rates are still bounded away from zero.
In Sections~\ref{sec:myopic-marc} and \ref{sec:myopic-brc}, we investigate myopic decode-forward for two other channels, namely the multiple-access relay channel and the broadcast relay channel. We show that under certain conditions, achievable rates of myopic decode-forward can be as large as that of omniscient decode-forward. We conclude the paper in Section~\ref{sec:myopic-conclusion}.

\section{Myopic Coding}\label{sec:myopic-coding}
\subsection{What is Myopic Coding?}
Recall that we categorize a coding strategy as omniscient if all nodes have a global view of the network and can cooperate completely. Now, we define myopic coding. 
This is an informal definition which will be made more precise later in the paper.

\begin{indef}
A myopic \texttt{X} coding strategy  is a constrained version of the corresponding omniscient \texttt{X}  coding strategy. The constraint in myopic coding is such that every node cooperates with only a few other nodes.  This cooperation can be in the form of transmitting to another node, processing (e.g., decoding, amplifying, quantizing) or canceling the transmissions from another node.
\end{indef}

We note that a myopic coding strategy is defined with respect to an omniscient coding strategy. Though there is no fixed way of constraining an omniscient coding strategy, the idea is to limit the processing at the nodes by limiting the number of neighbors a node communicates and cooperates with. Myopic coding aims to achieve practical advantages, e.g., lower computational complexity, robustness to topology changes, and fewer storage/buffer requirements.

To illustrate myopic coding, we now briefly discuss two myopic coding strategies for the multiple-relay channel, namely myopic decode-forward  and myopic amplify-forward.

\subsection{Myopic Decode-Forward for the Multiple-Relay Channel}
Let us consider the decode-forward coding strategy for the multiple-relay channel by Xie and Kumar~\cite{xiekumar03}, in which every message is fully decoded at and forwarded by the relays. It is also known as the \emph{decode-and-forward} strategy. In this strategy, block Markov encoding (irregular block Markov encoding\footnote{We use the terminology in \cite{kramergastpar04}. Note that the terms were not used in the original paper but subsequently used in later papers.} \cite{covergamal79} and regular block Markov encoding\footnotemark[\value{footnote}]\cite{willemsmeulen83}) can be used.  In the Gaussian channel, a node splits its total transmission power between sending new information and repeating what the relays \emph{in front} (downstream, i.e., toward the destination) send.  For decoding, successive decoding\footnotemark[\value{footnote}] \cite{covergamal79} can be used for irregular Markov encoding; backward decoding \cite{willemsmeulen85} or sliding window decoding\footnotemark[\value{footnote}] \cite{kramergastpar03} can be used for regular block Markov encoding.  In the Gaussian channel, a node decodes signals from all the nodes \emph{behind} (upstream, i.e., toward the source).  At the same time, it cancels interfering transmissions from all the nodes in front.  Since all the nodes fully cooperate, we term this coding strategy \emph{omniscient decode-forward}.

\begin{figure}[t]
\begin{minipage}[b]{0.48\linewidth}
\centering
\includegraphics[width=\textwidth]{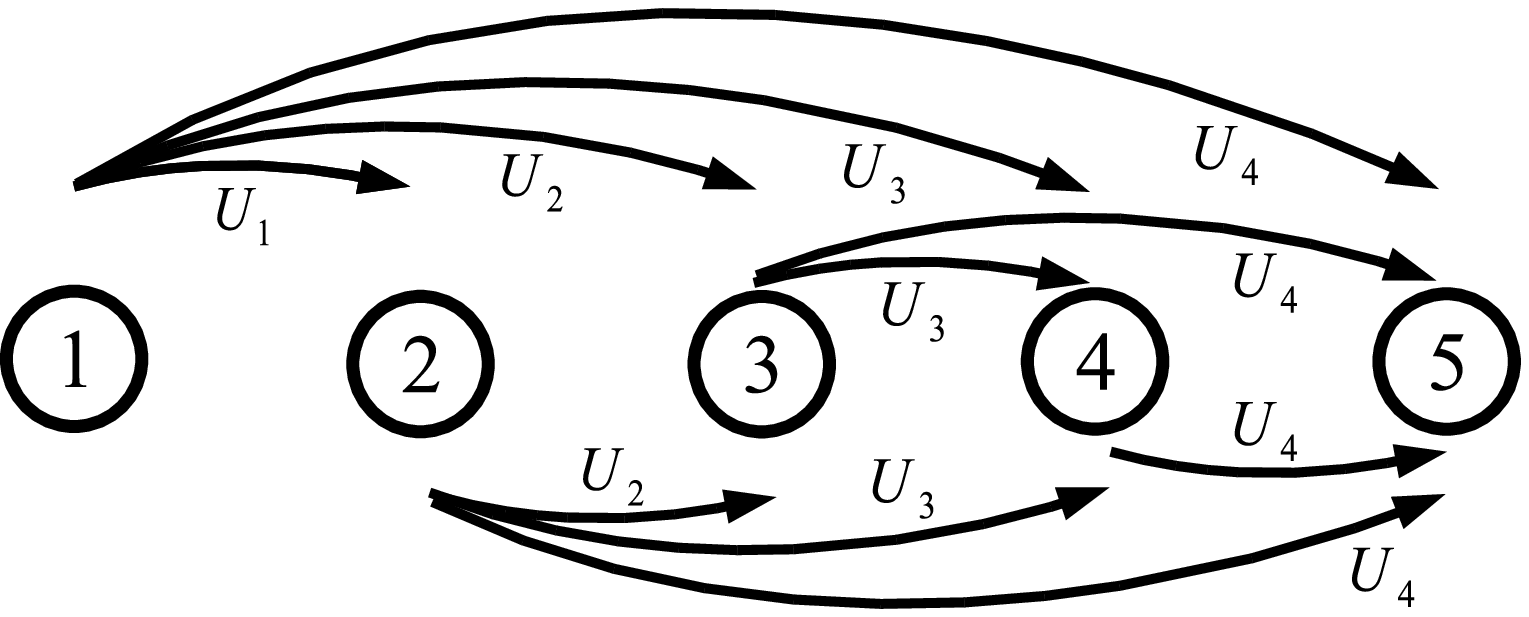}
\caption{Omniscient decode-forward for the five-node Gaussian multiple-relay channel.} \label{fig:5_node_mr_complete_view}
\end{minipage}
\hspace{0.3cm}
\begin{minipage}[b]{0.48\linewidth}
\centering
\includegraphics[width=\textwidth]{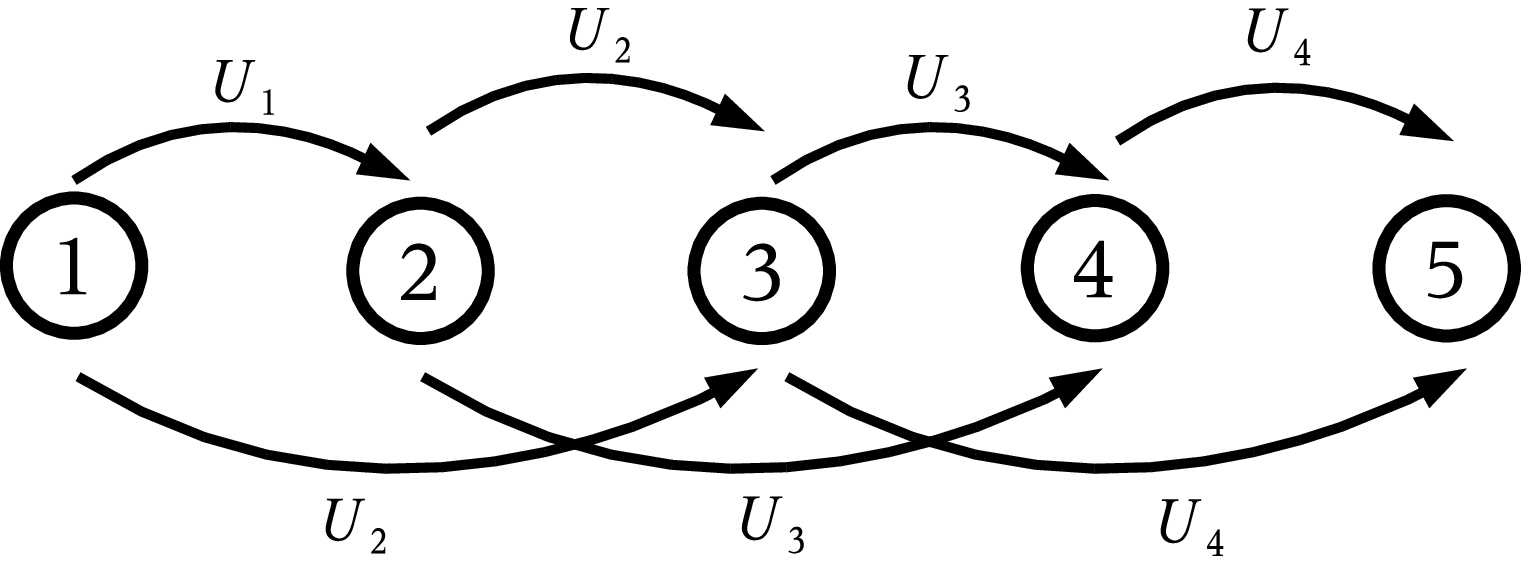}
\caption{Two-hop myopic decode-forward for the five-node Gaussian multiple-relay channel.} \label{fig:5_node_mr_2_hop}
\end{minipage}
\end{figure}

Now, we use an example to illustrate how each node cooperates with all other nodes in omniscient decode-forward. Consider a five-node Gaussian multiple-relay channel (the formal definition can be found in Section~\ref{sec:def-gaussian-mrc}). Using omniscient decode-forward, a node transmits to all the nodes in front. Fig.~\ref{fig:5_node_mr_complete_view} depicts the transmissions of the nodes. Let all $U_i, i=1,2,3,4$, be independent random variables. When node 4 transmits $U_4$ to node 5, node 3 splits its power, transmitting new information ($U_3$) to node 4 and helping node 4 to transmit another copy of what node 4 transmits ($U_4$) to node 5. Similarly, nodes 1--3 split their power to transmit new information and old information (the same information of what the nodes in front transmit). In decoding, a node decodes the transmissions from all nodes behind. For example, node 5 decodes all transmissions from nodes 1--4. In addition, a node cancels all transmissions from the nodes in front when it decodes. For example, when node 2 decodes $U_1$ from node 1, it cancels $U_3$ and $U_4$ from node 3, $U_4$ from node 4, as well as $U_2, U_3$, and $U_4$ from node 1.

Now, we consider a myopic version of the omniscient decode-forward in which nodes are limited in how much information they can store and process.  We define $k$-\emph{hop myopic decode-forward} for the multiple-relay channel as follows.
\begin{defn}
$k$-hop myopic decode-forward for the multiple-relay channel is a constrained version of omniscient decode-forward, and the constraints are as follows.
\begin{itemize}
\item In encoding, a node must transmit messages that it has decoded from at most the past $k$ blocks of received signal.
\item In decoding, a node can decode one message using only $k$ blocks of received signal.
\item A node can store a decoded message in its memory over at most $k$ blocks.
\end{itemize}
\end{defn}

At the first glance, the above constraints for myopic decode-forward do not seem to include the view of a node or how many other nodes a node can communicate with. However, these are embedded in the definition itself. The constraints automatically restrict the number of nodes a node can cooperate with. Furthermore, the restrictions stem from practical advantages of having fewer processing and storage requirements at the nodes, which are the motivations behind myopic coding.

Now, let us consider \emph{two-hop myopic decode-forward}. The encoding and the decoding processes at the nodes in the five-node multiple-relay channel are as follows (refer to Fig.~\ref{fig:5_node_mr_2_hop})
\begin{itemize}
\item Node 1 transmits $U_1$ and $U_2$, node 2 transmits $U_2$ and $U_3$, etc.
\item Node 5 decodes $U_3$ and $U_4$, node 4 decodes $U_2$ and $U_3$, etc.
\item During decoding, node 2 cancels $U_2$ and $U_3$, node 3 cancels $U_3$ and $U_4$, etc. 
\end{itemize}

We note that this encoding technique is different from \cite[Fig. 1]{guptakumar03}, in which the source and the relay transmit independent signals (hence no coherent combining is possible) while the relays and the destination decode transmissions from all nodes behind. The decoding technique in \cite{guptakumar03} is only possible under omniscient coding as a node decodes each message using the received signals from all upstream nodes, possibly over a large number of blocks. 

In myopic decode-forward for the multiple-relay channel, we use the concept of regular block Markov encoding and sliding window decoding.  However, the encoding and the decoding techniques differ from that found in the literature as the nodes have limited views.  It is noted that myopic coding captures point-to-point coding and omniscient coding as special cases.
In particular, $k$-hop myopic decode-forward for the multiple-relay channel where $k=1$ is point-to-point coding  and $k=T-1$ ($T$ is the number of nodes in the channel) omniscient decode-forward.

The reader is reminded that the term ``hop'' used here does not carry the same meaning as it does in multi-hop routing. The term hop is best understood by looking at the sequence in which the messages are decoded, e.g., if the messages are decoded by node $i$ followed by node $j$, then node $j$ is node $i$'s next hop.

We say that a set of nodes $\mathcal{V}$ are in the \emph{view} of node $i$ if node $i$ processes (e.g., decodes, amplifies, or quantizes) or cancels the transmissions from all the nodes in $\mathcal{V}$.

\subsection{Myopic Amplify-Forward for the Multiple-Relay Channel}
Next, let us consider the amplify-forward strategy for the multiple-relay channel by Yuksel and Erkip~\cite{yukselerkip03}. We will use the one-source, two-relay, one-destination network as an example. Consider the ``$S+R_1(S)+R_2(S,R_1)$'' scheme \cite[Table I]{yukselerkip03}. In this scheme, the transmissions are split into three blocks. In block 1, the source transmits to both relays and the destination (hence the notation $S$). In block 2, relay 1 normalizes its received signal from the source in block 1 and forwards the normalized received signal to relay 2 and the destination (hence the notation $R_1(S)$). Relay 2 combines the signals that it has received in blocks 1 and 2, normalizes to its own power value, and transmits the combined signal in block 3 (hence the notation $R_2(S,R_1)$). The destination then decodes using the three blocks of received signal (hence the notation $S+R_1(S)+R_2(S,R_1)$). We term this coding strategy omniscient amplify-forward, as each node cooperates with all other nodes.

Now, let us consider a myopic version of the amplify-forward strategy. It has been noted in \cite{yukselerkip03} that relay 2 can choose to listen to only relay 1 (which transmits in block 2) and forwards only this received signal to the destination (the notation used is $R_2(R_1)$). Instead of decoding over three blocks, the destination can choose to decode only from relay 2 (which transmits in block 3). We see that in this scheme, a node listens to only one node and forwards to another node. Hence, we term this strategy \emph{one-hop} myopic amplify-forward.  One can similarly construct two-hop myopic amplify-forward, and so on.

\subsection{Practical Advantages of Myopic Coding}
In this section, we discuss a few practical advantages of myopic coding compared to omniscient coding. These include simpler code design, increased robustness, reduced computation and memory requirements, and local synchronization.
Though the analyses of myopic coding in this paper are based on information-theoretic achievable rates (in Shannon's sense), the practical advantages here are relevant to code designs based on these strategies (myopic or omniscient, decode-forward or amplify-forward, etc.).
That researchers are interested in practical implementations of information-theoretic cooperative strategies is apparent in the recent work that has been proposed in this direction.
There are various codes designed based on omniscient decode-forward for the single-relay channel~\cite{razaghiyu06,ezrigastpar06,khojastepouraazhang04,chakrabarti06} and the multiple-relay channel~\cite{yu06,ongmotani07asecon,ongmotani07bisit}.
One may design myopic versions of these codes to reap the practical advantages discussed in this section. 

Looking closely at the LDPC codes using parity forwarding (based on omniscient decode-forward) for the multiple-relay channel \cite{yu06}, we see that the complexity of designing codes grows with the number of relays. This means that constructing codes in which all nodes cooperate can be more difficult compared to designing codes in which nodes only cooperate with neighboring nodes. This technique of utilizing local knowledge (or limited cooperation) is prevalent in other wireless network problems, e.g., cluster-based routing~\cite{jiangli99}, whereby nodes are split into clusters, and routes are optimized locally.

Myopic coding schemes are more robust to topology changes than the corresponding omniscient coding schemes. For example, consider cancellation of the interference from downstream nodes. In omniscient coding, a node needs to have the knowledge or an estimate of what every downstream node transmits in order to cancel it. Any error in the cancellation (due to topology changes or node failures not known to the decoder) will affect the decoding and thus the rate. 
In myopic coding, nodes only cancel the interference from a few neighboring nodes. This means that topology changes or node failures beyond a node's view  are less likely to affect its decoding. In Appendix~\ref{myopic-robustness-example}, we give another example to show how node failures affect more nodes in myopic coding than in omniscient coding.

In addition, the encoding and decoding computations at each node under myopic coding can be less. Since a node only needs to transmit to and decode from a few nodes, the node encodes fewer data for its transmissions and decodes fewer data from the received signals.

Furthermore, since the nodes need to buffer fewer data for encoding, interference cancellation, and decoding, less memory is required for buffering and codebook storage. Consider the five-node Gaussian multiple-relay channel. Using omniscient decode-forward, node 1 encodes a message four times over four blocks, using different power splits. Node 5 buffers four blocks of its received signal to decode one message. The buffer grows as the number of nodes in the network increases. On the other hand, using myopic decode-forward, the nodes buffer fewer blocks of received signal, and the buffer size for each node is independent of the number of nodes in the network.

Myopic coding mitigates the need for synchronization of the entire network. Under omniscient decode-forward, all the nodes might need to be synchronized. On the other hand, under myopic coding, a node only needs to synchronize with a few neighboring nodes. Hence, synchronization can be done locally.

In brief, myopic coding can increase the robustness and scalability of the network. In the next section, we analyze the performance of myopic coding in the multiple-relay channel using the decode-forward coding strategy.

\section{Myopic Coding in the Multiple Relay Channel}\label{sec:myopic-mrc}
In this section, we construct random codes for myopic decode-forward for the multiple-relay channel and compare the performance of these myopic coding strategies to the corresponding omniscient coding strategy.

\subsection{Channel Model}\label{sec:model}
\begin{figure}[t]
\centering
\includegraphics[width=11cm]{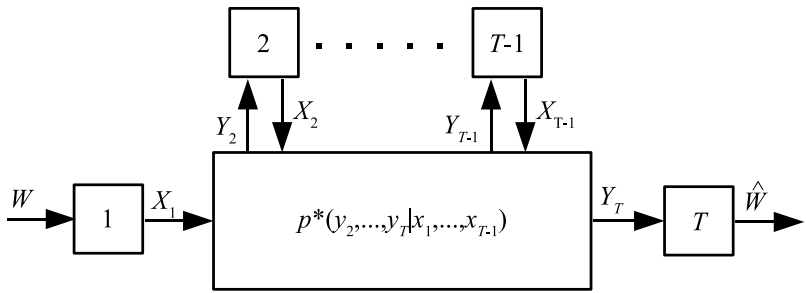}
\caption{The $T$-node multiple-relay channel.} \label{fig:1DSensor}
\end{figure}

Fig.~\ref{fig:1DSensor} depicts the $T$-node multiple-relay channel, with node 1 being the source and node $T$ the destination.  Nodes 2 to $T-1$ are purely relays. Message $W$ is generated at node 1 and is to be sent to node $T$.  A multiple-relay channel can be completely described by the channel distribution
\begin{equation}
p^*(y_2, y_3, \dotsc, y_T | x_1, x_2, \dotsc, x_{T-1})
\end{equation}
on $\mathcal{Y}_2 \times \mathcal{Y}_3 \times \dotsm \times \mathcal{Y}_T$, for each $(x_1, x_2, \dotsc, x_{T-1}) \in \mathcal{X}_1 \times \mathcal{X}_2 \times \dotsm \times \mathcal{X}_{T-1}$.  In this paper, we only consider memoryless and time invariant channels \cite{kramergastpar04}, which means
\begin{equation}
p(y_{2i}, \dotsc, y_{Ti} | x^i_1, \dotsc, x^i_{T-1}, y^{i-1}_2, \dotsc, y^{i-1}_T) = p^*(y_{2i}, \dotsc, y_{Ti} | x_{1i}, \dotsc, x_{(T-1)i}),
\end{equation}
for all $i$.
We use the following notation: $x_i$ denotes an input from node $i$ into the channel; $x_{ij}$ denotes the $j$-th input from node $i$ into the channel; $y_{ij}$ denotes the $j$-th output from the channel to node $i$; and $x^i_t = x_{t1}, x_{t2}, \dotsc, x_{ti}$.

We denote the $T$-node multiple-relay channel by the tuple
\begin{equation}
\Big( \mathcal{X}_1 \times \dotsm \times \mathcal{X}_{T-1}, p^*(y_2, \dotsc, y_T | x_1, \dotsc, x_{T-1}), \mathcal{Y}_2 \times \dotsm \times \mathcal{Y}_T \Big).
\end{equation}

\subsection{Notation and Definitions}
In the multiple-relay channel, the information source at node 1 emits
random letters $W$, each taking on values from a finite set of size
$M$, that is $w \in \{ 1, ..., M \} \triangleq \mathcal{W}$. We consider each $n$ uses of the channel as a block.

\begin{defn}
An $( M, n )$ code of a $T$-node multiple-relay channel comprises:
\begin{itemize}
\item An encoding function at node 1, $f_{1} : \mathcal{W} \rightarrow \mathcal{X}_1^n$, which maps a source letter to a codeword of length $n$.
\item $n$ encoding functions at node $t, t=2, 3, \dotsc, T-1$, $f_{ti}: \mathcal{Y}_t^{i-1} \rightarrow \mathcal{X}_t, i=1, 2, \dotsc, n$, such that $x_{ti} = f_{ti}(y_{t1}, y_{t2}, \dotsc, y_{t(i-1)})$, which map past received signals to the signal to be transmitted into the channel.
\item A decoding function at the destination, $g_{T}: \mathcal{Y}_T^{n} \rightarrow \mathcal{W}$, such that $\hat{w}=g_{T}(y_T^n)$,
which maps received signals of length $n$ to a source letter estimate.
\end{itemize}
\end{defn}

\begin{defn}
Assuming that the source letter $W$ is uniformly
distributed over $\{ 1, ..., M \}$, the average error probability is defined
as
\begin{equation}
P_e = \Pr \{\hat{W} \neq  W \}.
\end{equation}
\end{defn}
We denote the estimated $i$-th source letter at the destination as $\hat{W}_i$.

\begin{defn} \label{def:ratePair}
The rate
\begin{equation} \label{eqn:ratepair}
R \leq \frac{1}{n} \log M
\end{equation}
is achievable if, for any $\epsilon > 0$, there is at least one $(M,n)$ code such that $P_e < \epsilon$.
\end{defn}

The following definition and lemma are taken from \cite[p. 384]{coverthomas91} and \cite[p. 386]{coverthomas91} respectively.
\begin{defn}\label{def:AEP}
Consider a finite collection of random variables
$(X_1,X_2,\dotsc,X_k)$ with some fixed joint distribution
$p(x_1,x_2,\dotsc,x_k)$. Let $S$ denote an arbitrarily ordered subset of these
random variables, and consider $n$ independent copies of $S$.
\begin{equation}
\Pr\{ \mathbf{S}=\mathbf{s} \} = \prod_{i=1}^n \Pr \{ S_i = s_i \}.
\end{equation}
The set $\mathcal{A}_\epsilon^n$ of $\epsilon$-typical $n$-sequences
$(\mathbf{x}_1,\mathbf{x}_2,\dotsc,\mathbf{x}_k)$ is defined as
\begin{equation}
\mathcal{A}_\epsilon^n(X_1,X_2,\dotsc,X_k) = \bigg\{
(\mathbf{x}_1,\mathbf{x}_2,\dotsc,\mathbf{x}_k) : \left|
-\frac{1}{n} \log p(\mathbf{s}) - H(S) \right| < \epsilon , \quad \forall
S \subseteq \{X_1,X_2,\dotsc,X_k\} \bigg\}.
\end{equation}
\end{defn}

\begin{lem}\label{lem:AEP}
For any $\epsilon > 0$ and for sufficiently large $n$, $\left|
\mathcal{A}_\epsilon^n(S) \right| \leq
    2^{n(H(S)+\epsilon)}$
\end{lem}

Throughout this paper, we follow the notation for node permutation used in \cite{kramergastpar03}. Let $\mathcal{T}$ be the set of all
relay nodes, $\mathcal{T} = \{ 2, 3, \dotsc, T-1 \}$.  Let
$\pi(\cdot)$ be a permutation on $\mathcal{T}$.  Define $\pi(1) =
1$, $\pi(T) = T$ and $\pi(i:t) = \{ \pi(i),  \pi(i+1),  \dotsc,  \pi(t)
\}$.

\subsection{The Gaussian Multiple-Relay Channel} \label{sec:def-gaussian-mrc}
In the $T$-node Gaussian multiple-relay channel, node $t$, $t=2, \dotsc, T$, receives
\begin{equation}
Y_t = \sum_{\substack{i=1, \dotsc, T-1 \\ i \neq t}} \sqrt{\lambda_{it}}X_i + Z_t,
\end{equation}
where $X_i$, input to the channel form node $i$, is a random variable with fixed average power $E[X_i^2] = P_i$. $Y_t$ is the received signal at node $t$. $Z_t$, the receiver noise at node $t$, is an independent zero-mean Gaussian random variable with variance $N_t$. $\lambda_{it}$ is the channel gain from node $i$ to node $t$. $\lambda_{it}$ depends on the antenna gain, the carrier frequency of the transmission, and the distance between the transmitter and the receiver.

We consider Gaussian multiple-relay channels with fixed average transmit power at the source and at all relays. We note that using omniscient decode-forward, having a maximum average power constraint on every node is equivalent to having a fixed average transmit power constraint on the node, as the overall rate is a non-decreasing function of the average transmit power at any node, keeping the transmit power of other nodes constant. This is because a node decodes the transmissions from all upstream nodes and cancels the transmissions from all downstream nodes. So, the transmissions of all nodes are either used in decoding or canceled but are never treated as noise. However, under myopic coding, lowering the transmit power at certain nodes may help to reduce the interference at other nodes and increase the overall rate. Hence the maximum rate achievable by myopic decode-forward with maximum average power constraints on the nodes is lower bounded by that with fixed average power constraints.

We use the standard path loss model for signal propagation. The channel gain is given by
\begin{equation}
\lambda_{it} = \kappa d_{it}^{-\eta},
\end{equation}
where $\eta$ is the path loss exponent, and $\eta \geq 2$ with equality for free space transmission. $\kappa$ is a positive constant as far as the analyses in this paper are concerned. Hence, the received power at node $t$ from node $i$ is given by
\begin{equation}\label{eq:path_loss}
P_{it} = \lambda_{it}P_i = \kappa d_{it}^{-\eta}P_i.
\end{equation}

For the channel where all transmitters have the same power constraint, i.e., $P_i = P$, and all receivers have the same noise power, i.e., $N_t = N$, we define the signal-to-noise ratio (SNR) to be $\frac{P}{N}$.

\subsection{Achievable Rates}\label{sec:achievable_rate}

In this section, we investigate achievable rates of two myopic decode-forward coding strategies and the omniscient decode-forward coding strategy.

\subsubsection{Omniscient coding}
First, we consider achievable rates of omniscient decode-forward. Xie and Kumar~\cite{xiekumar03} proposed a decode-forward coding strategy for the multiple-relay channel. They showed that the following rate is achievable,
which is higher than that in \cite{guptakumar03}.
\begin{subequations}
\begin{align}
R & \leq \max_{\pi(\cdot)} \max_{p(\cdot)} \min_{
1\leq t \leq T-1} I(X_{\pi(1:t)} ; Y_{\pi(t+1)} | X_{\pi(t+1;T-1)}) \\
&= R_{\text{omniscient}}.
\end{align}
\end{subequations}

The first maximization allows us to arrange the order in which data flow through the relay nodes.  The second maximization
is over all possible distributions $p(x_1,x_2,\dotsc,x_{T-1})$ on $\mathcal{X}_1 \times \dotsm \times \mathcal{X}_{T-1}$.  The
minimization is over all relays and the destination, where full decoding of the messages must be done. Since
all the information must pass through each relay, the relay that decodes at the lowest rate becomes the bottleneck of the overall
transmission.  We note in the mutual information term that
node $\pi(t+1)$ receives the transmission from all nodes
behind, $X_{\pi(1:t)}$. Since it knows what the nodes
in front transmit (by the flow of data), it can cancel
out their transmissions, as seen in the conditioned term
$X_{\pi(t+1;T-1)}$.

Now, we investigate achievable rates of myopic decode-forward coding strategies. We note that using decode-forward, all relays must fully decode the messages. We assume that the relays decode the messages sequentially.

\subsubsection{One-Hop Myopic Coding (Point-to-Point Coding)}
In one-hop myopic decode-forward, a relay node transmits what it has decoded from one block of received signal. This means a node transmits to only the node in the next hop.  In decoding, a node decodes one message using one block of received signal. This means a node decodes from only one node behind. A node keeps its decoded message for one block, and it uses the last decoded message to cancel the effect of its own transmission. Using random coding \cite{shannon48}, node $\pi(t)$ can reliably decode data up to the rate
\begin{equation}\label{eq:recep_rate_1hop}
R_{\pi(t)} = I(X_{\pi(t-1)};Y_{\pi(t)}|X_{\pi(t)}),
\end{equation}
for some $p(x_1)p(x_2)\dotsm p(x_{T-1})$, $t \in \{ 2, \dotsc, T\}$, and $X_{\pi(T)}=0$.  Since all information must pass through all nodes in order to reach the
destination, the overall rate is constrained by
\begin{equation}
R \leq \min_{t \in \{ 2, \dotsc, T\}} R_{\pi(t)}.
\end{equation}
Noting that the messages can flow through the relays in any order \cite{kramergastpar03} and the nodes transmit independent signals, we have the following result.

\begin{thm}
Let
\begin{equation}
\Big( \mathcal{X}_1 \times \dotsm \times \mathcal{X}_{T-1}, p^*(y_2, \dotsc, y_T | x_1, \dotsc, x_{T-1}), \mathcal{Y}_2 \times \dotsm \times \mathcal{Y}_T \Big)\nonumber
\end{equation}
be a memoryless multiple-relay channel.  Under one-hop myopic decode-forward or point-to-point coding, the rate $R$ is achievable, where
\begin{equation}
R \leq \max_{\pi(\cdot)} \max_{p(\cdot)} \min_{t \in \{2, \dotsc, T\}} I(X_{\pi(t-1)};Y_{\pi(t)}|X_{\pi(t)}) = R_{\text{1-hop}}.
\end{equation}
The outer maximization is over all possible node permutations and the inner maximization is taken over all joint distributions of the form
\begin{equation}
p(x_1, \dotsc, x_{T-1}, y_2, \dotsc, y_T) = p(x_1)p(x_2)\dotsm p(x_{T-1})
\times p^*(y_2, \dotsc, y_T | x_1, \dotsc, x_{T-1})\nonumber.
\end{equation}
\end{thm}

\subsubsection{Two-Hop Myopic Coding}
Instead of just transmitting to only its immediate neighbor, a node might want to help the neighboring node to transmit to the neighbor's neighbor. Under two-hop myopic decode-forward, a node can transmit messages that it has decoded in the past two blocks of received signals. That means in block $i$, a node transmits data that it has decoded in blocks $i-1$ and
$i-2$. In decoding, it decodes one message using only two blocks of received signal.
Two-hop myopic decode-forward achieves rates up to that given in the following theorem.

\begin{thm}\label{thm:two-hop_myopic}
Let
\begin{equation}
\Big( \mathcal{X}_1 \times \dotsm \times \mathcal{X}_{T-1}, p^*(y_2, \dotsc, y_T | x_1, \dotsc, x_{T-1}),
\mathcal{Y}_2 \times \dotsm \times \mathcal{Y}_T \Big)\nonumber
\end{equation}
be a $T$-node memoryless multiple-relay channel.  Using two-hop myopic decode-forward, the rate $R$ is achievable, where
\begin{subequations}
\begin{align}
R &\leq \max_{\pi(\cdot)} \max_{p(\cdot)} \min_{t \in \{2, \dotsc, T\}} I(U_{\pi(t-2)},U_{\pi(t-1)};Y_{\pi(t)}|U_{\pi(t)},U_{\pi(t+1)})\\
&= R_{\text{2-hop}},
\end{align}
\end{subequations}
where $U_{\pi(0)} = U_{\pi(T)} = U_{\pi(T+1)} = 0$, for $\pi(0)=0$ and $\pi(T+1)=T+1$. The outer maximization is over all possible relay permutations and the inner maximization is taken over all joint distributions of the form
\begin{subequations}
\begin{align}
& p(x_1, x_2 \dotsc, x_{T-1}, u_1, u_2 \dotsc, u_{T-1}, y_2, y_3 \dotsc, y_T) \nonumber\\
& = p(u_{\pi(1)})p(u_{\pi(2)})\dotsm p(u_{\pi(T-1)}) p(x_{\pi(1)}|u_{\pi(1)},u_{\pi(2)})   p(x_{\pi(2)}|u_{\pi(2)},u_{\pi(3)}) \dotsm p(x_{\pi(T-1)}|u_{\pi(T-1)})\nonumber\\
& \quad \times p^*(y_2, \dotsc, y_T | x_1, \dotsc, x_{T-1}). \nonumber
\end{align}
\end{subequations}
\end{thm}

The proof of Theorem~\ref{thm:two-hop_myopic} can be found in Appendix~\ref{append:two-hop}.

Using a particular probability distribution function on a coding strategy, we term the maximum rate at which a node can reliably decode the source messages the \emph{reception rate}. For example, using one-hop myopic decode-forward, the reception rate at node $\pi(t)$ is $R_{\pi(t)} = I(X_{\pi(t-1)};Y_{\pi(t)}|X_{\pi(t)})$; using two-hop myopic decode-forward, the reception rate at node $\pi(t)$ is $R_{\pi(t)} = I(U_{\pi(t-2)},U_{\pi(t-1)};Y_{\pi(t)}|U_{\pi(t)},U_{\pi(t+1)})$.

\subsection{Performance Comparison}\label{sec:performance}

In this section, we compare achievable rates of the two myopic coding strategies and the omniscient coding strategy for the Gaussian multiple-relay channel.

\subsubsection{Channel Setup}
Consider a linear five-node channel, in which nodes are arranged in a straight line in the sense that for any $i < j < k, d_{ik} = d_{ij} + d_{jk}$. Node 1 is the
source, nodes 2, 3, and 4 are the relays, and node 5 is the
destination.  Node $t$, $t = 2, 3, 4, 5$, receives the following channel output,
\begin{equation}
Y_t = \sum_{\substack{i=1\\i\neq t}}^4 \sqrt{\kappa d_{it}^{-\eta}}X_i
+ Z_t.
\end{equation}
In all analyses in this section, we use the following parameters:
$N_2 = N_3 = N_4 = N_5 = N = 1$W, $\kappa = 1$, and $\eta = 2$.

Now, consider a point-to-point link. The rate at which information
can be transmitted through a Gaussian channel (per channel use) from
node $i$ to node $t$ is given by~\cite{coverthomas91}
\begin{equation}\label{eq:gaussian_rate}
R \leq \frac{1}{2} \log \left( 1 + \frac{P_{it}}{N_t} \right).
\end{equation}
Throughout this paper, logarithm base 2 is used and hence the units of rate are bits per channel use.

\subsubsection{One-Hop Myopic Coding}

In one-hop myopic decode-forward, node $t$ transmits only to node $t+1$.  Let
us first consider node 1.  It sends $X_1$ to node 2. Node
2 receives
\begin{equation}
Y_2 = \sqrt{\kappa d_{12}^{-\eta}}X_1 +  \sqrt{\kappa d_{32}^{-\eta}}X_3 +
\sqrt{\kappa d_{42}^{-\eta}}X_4 + Z_2.
\end{equation}
Node 2 decodes new messages from node 1's transmission. From \eqref{eq:recep_rate_1hop}, the reception rate at node 2 is
\begin{subequations}
\begin{align}
R_2 &= I(X_1;Y_2|X_2)\\
& = \frac{1}{2} \log 2\pi e \left[\kappa d_{12}^{-\eta}P_1 + \kappa d_{23}^{-\eta}P_3 + \kappa d_{24}^{-\eta}P_4 + N_2 \right] - \frac{1}{2} \log 2\pi e
\left[ \kappa d_{23}^{-\eta}P_3 + \kappa d_{24}^{-\eta}P_4 + N_2 \right]\\
& = \frac{1}{2} \log \left[ 1 +
\frac{d_{12}^{-2}P_1}{1 + d_{23}^{-2}P_3 + d_{24}^{-2}P_4 } \right].
\end{align}
\end{subequations}
Here, we have substituted $\kappa =1$, $\eta = 2$, and
$N_2=1$W. The reception rates at nodes 3, 4, and 5 can be computed in similar way. Achievable rates of one-hop myopic decode-forward are
\begin{equation}
R \leq \min_{t \in \{2,3,4,5\}} R_t = R_{\text{1-hop}}.
\end{equation}
We note that the message flow through the nodes in the order $\{1,2,3,4,5\}$ gives the highest achievable rate in this network.

\begin{figure}[t]
\begin{minipage}[b]{0.48\linewidth}
\centering
\includegraphics[width=\textwidth]{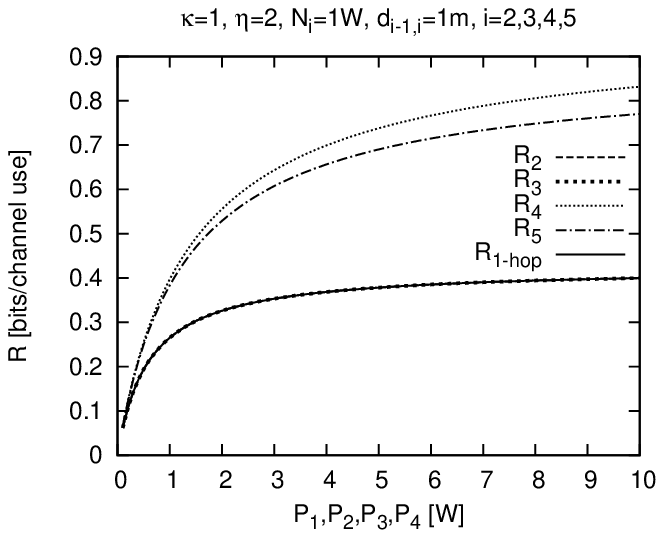}
\caption{Achievable rates of one-hop myopic decode-forward for the five-node multiple-relay channel, with equal node spacing.} \label{fig:rate_power_1hop_1}
\end{minipage}
\hspace{0.3cm}
\begin{minipage}[b]{0.48\linewidth}
\centering
\includegraphics[width=\textwidth]{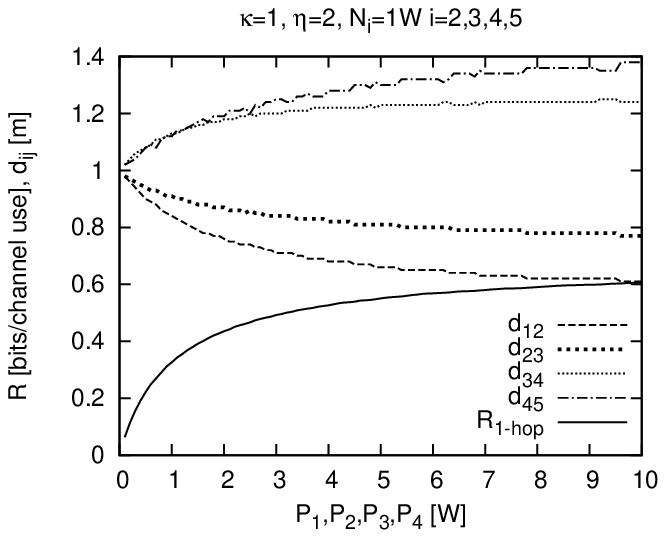}
\caption{Achievable rates of one-hop myopic decode-forward for the five-node
multiple-relay channel, with the optimal node spacing.} \label{fig:rate_power_1hop_2}
\end{minipage}
\end{figure}

Figs.~\ref{fig:rate_power_1hop_1} and \ref{fig:rate_power_1hop_2} show achievable rates of one-hop myopic decode-forward for equal node spacing and the optimal node spacing respectively.  In the latter, the spacing among the nodes is determined by brute force, with the constraints that all five nodes form a straight line (node $i+1$ is in front of node $i$) and $d_{15}=4$.

When the nodes are equally spaced, $R_{\text{1-hop}}$ is constrained by reception rates $R_2$ and $R_3$.  In order to increase $R_2$ and $R_3$, the distance $d_{12}$ and $d_{23}$ should be decreased.  We see that this is indeed the case.  The optimum values for $d_{12}$ and $d_{23}$ are less than 1m, as can be seen in Fig.~\ref{fig:rate_power_1hop_2}.

We see in Fig.~\ref{fig:rate_power_1hop_2} that as the average transmit power increases, the optimal $d_{12}$ and $d_{23}$ decrease while the optimal $d_{34}$ and $d_{45}$ increase. This is because $R_2$ and $R_3$ are significantly affected when $P_3$ and $P_4$ increase. Recall that in one-hop myopic decode-forward, a node treats the transmissions of all the nodes beyond its view as noise. For example, node 3 decodes from node 2, and treats the transmissions of nodes 1 and 4 as noise. Since there is no transmitting node in front of node 4, $R_4$ and $R_5$ are less affected by the increase of the transmit power. Hence, to compensate for the greater noise experienced by nodes 2 and 3 as the transmit power increases, $d_{12}$ and $d_{23}$ are reduced to increase $R_2$ and $R_3$.

\subsubsection{Two-Hop Myopic Coding}
In two-hop myopic decode-forward, node $t, t = 1, 2, 3$, allocate $\alpha_t$
of its power to transmit to node $t+2$ and $(1-\alpha_t)$ of its
power to node $t+1$.  Since there is only one node in front
of node 4, it allocates all its power to transmit to node 5.  The transmission by each
node is listed as follows:
\begin{itemize}
\item Node 4 sends $X_4 = \sqrt{P_4}U_4$.
\item Node 3 sends $X_3 = \sqrt{\alpha_3P_3}U_4 + \sqrt{(1-\alpha_3)P_3}U_3$.
\item Node 2 sends $X_2 = \sqrt{\alpha_2P_2}U_3 + \sqrt{(1-\alpha_2)P_2}U_2$.
\item Node 1 sends $X_1 = \sqrt{\alpha_1P_1}U_2 + \sqrt{(1-\alpha_1)P_1}U_1$.
\end{itemize}
Here, $U_i, i=1, 2, 3, 4$ are independent Gaussian random variables, each with unit variance, $0 \leq \alpha_j \leq 1$ for $j=1,2,3$.

From \eqref{eq:recep_rate_2hop}, for fixed $\{\alpha_1, \alpha_2, \alpha_3\}$, the reception rate at node 2 is
\begin{subequations}
\begin{align}
R_2 & = I(U_1;Y_2|U_2,U_3)\\
& = \frac{1}{2} \log 2\pi e \Bigg[ \kappa d_{12}^{-\eta}(1-\alpha_1)P_1   + \left(\sqrt{\kappa d_{23}^{-\eta}\alpha_3P_3}+\sqrt{\kappa d_{24}^{-\eta}P_4}\right)^2
+ N_2 \Bigg]\nonumber\\
& \quad - \frac{1}{2} \log 2\pi e
\left[\left(\sqrt{\kappa d_{23}^{-\eta}\alpha_3P_3}
+\sqrt{\kappa d_{24}^{-\eta}P_4}\right)^2 + N_2 \right]\\
& = \frac{1}{2} \log \left[ 1 +
\frac{d_{12}^{-2}(1-\alpha_1)P_1}{1+\left(\sqrt{d_{23}^{-2}\alpha_3P_3}
+ \sqrt{d_{24}^{-2}P_4}\right)^2} \right].
\end{align}
\end{subequations}
Here, we have substituted $\kappa =1$, $\eta = 2$, and  $N_2 = 1$W. The reception rates at nodes 3, 4, and 5 can be computed in a similar way.

\begin{figure*}[t]
\begin{minipage}[t]{0.45\linewidth}
\centering
\includegraphics[width=\textwidth]{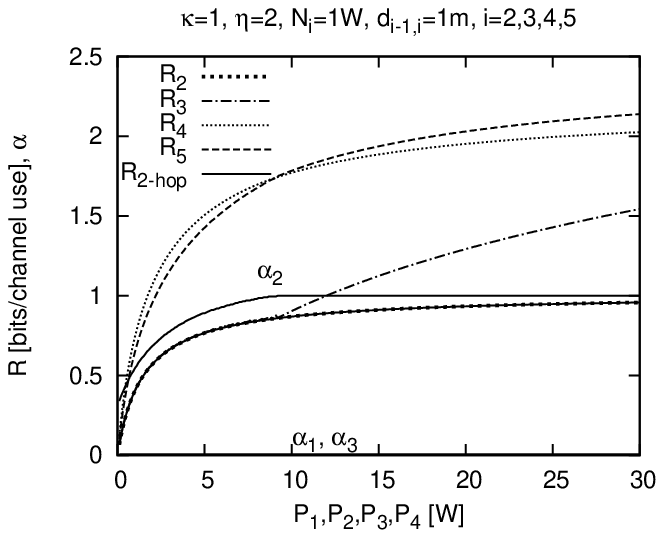}
\caption{Achievable rates of two-hop myopic decode-forward for the five-node
multiple-relay channel, with equal node spacing.}
\label{fig:rate_power_2hop_1}
\end{minipage}
\hfill
\begin{minipage}[t]{0.45\linewidth}
\centering
\includegraphics[width=\textwidth]{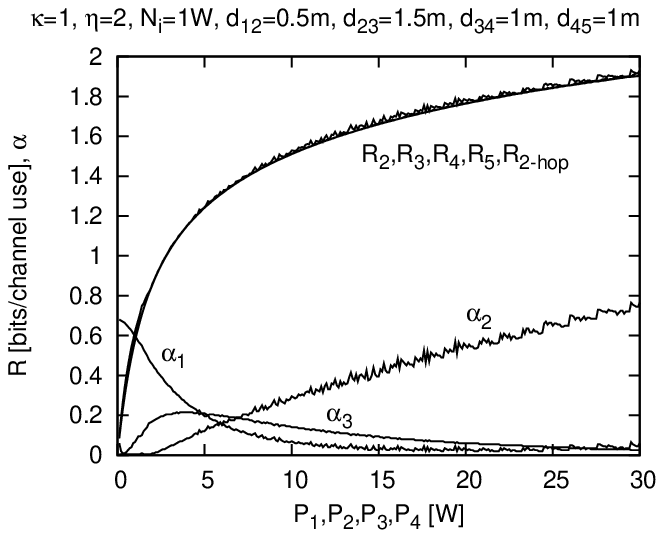}
\caption{Achievable rates of two-hop myopic decode-forward for the five-node
multiple-relay channel, with node 2 closer to the source.}
\label{fig:rate_power_2hop_2}
\end{minipage}
\begin{minipage}[b]{0.45\linewidth}
\centering
\includegraphics[width=\textwidth]{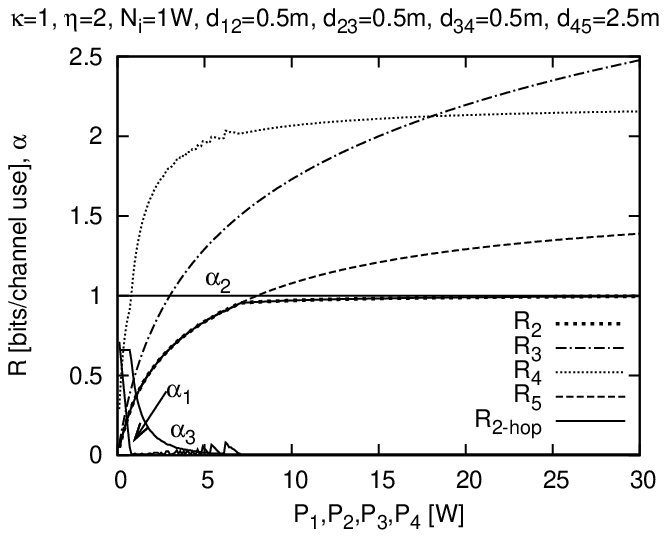}
\caption{Achievable rates of two-hop myopic decode-forward for the five-node
multiple-relay channel, with the relays clustered at the source.}
\label{fig:rate_power_2hop_3}
\end{minipage}
\hfill
\begin{minipage}[b]{0.45\linewidth}
\centering
\includegraphics[width=\textwidth]{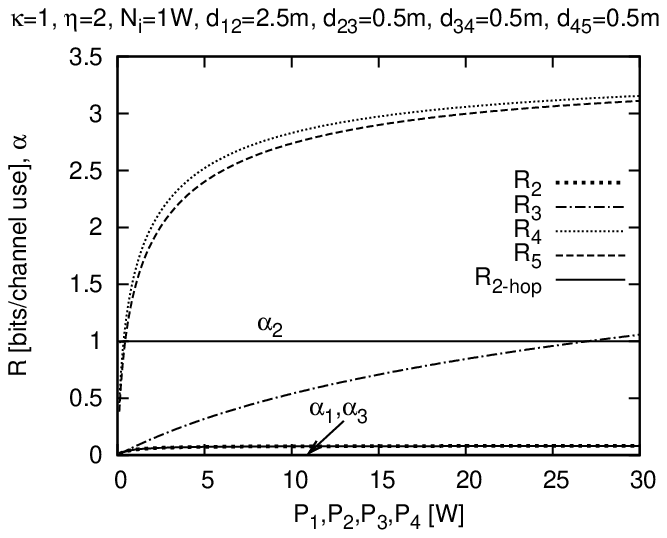}
\caption{Achievable rates of two-hop myopic decode-forward for the five-node
multiple-relay channel, with the relays clustered at the destination.}
\label{fig:rate_power_2hop_4}
\end{minipage}
\end{figure*}

Minimizing over all reception rates and maximizing over all possible power splits, the overall achievable rate is given by
\begin{equation}\label{eq:rate_2hop_view}
R \leq \max_{\{ \alpha_1,\alpha_2,\alpha_3 \}} \min_{t \in \{2,3,4,5\}} R_t = R_{\text{2-hop}}.
\end{equation}
We note that the message flow in the node permutation $\{1,2,3,4,5\}$ gives the highest overall rate in this network. Figs.~\ref{fig:rate_power_2hop_1}--\ref{fig:rate_power_2hop_4} show achievable rates, reception rates and power splits for nodes in different positions. We note that the nodes are arranged in a straight line.

When the nodes are equally spaced, we see that the overall rate is constrained by $R_2$ and $R_3$. Increasing the transmit power increases $R_3$ more than $R_2$. So, to maximize $\min \{R_2,R_3\}$, the optimal $\alpha_2$ increases to increase $R_2$ further. When the transmit power increases beyond 10W, $\alpha_2$ reaches it maximum and the overall rate is now restricted by $R_2$ alone. To understand this, we look at the rate equations.  For nodes 3--5, they decode the transmissions from 2 1/2 nodes behind, but node 2 decodes only from node 1. This makes $R_2$ the bottleneck of the overall transmission rate.  High $R_4$ and $R_5$ suggests that the overall rate can be improved by readjusting the position of the nodes.

One way to improve $R_2$ is to decrease $d_{12}$.  By doing this, we reduce the signal attenuation from
node 1 to node 2.  This indeed increases the overall rate, as shown
in Fig.~\ref{fig:rate_power_2hop_2}.  Here $d_{12}=0.5$m, while keeping the positions of
nodes 3, 4, and 5 unchanged.  Now, we see that the overall rate is constrained by $R_2, R_3, R_4$, and $R_5$, i.e., no single bottle-neck. We have seen that the increase in transmit power increases the reception rates of different nodes by different amount. Hence when the transmit power increases, the $\alpha$'s adjust themselves to maximize $\min \{R_2,R_3,R_4,R_5\}$.

Now, we study the cases when the relay nodes are clustered at the
source or at the destination.  Fig.~\ref{fig:rate_power_2hop_3} shows
achievable rates when the relays are clustered at the source.  In this
arrangement, the overall rate is constrained by both $R_2$ and $R_5$ when
the nodes transmit at low power,  and by $R_5$ alone when the nodes transmit at
high power.  That $R_5$ being the bottleneck
should not come as a surprise as node 5 is positioned far away from the
rest of the nodes.  However, at high power, the constraint is at $R_2$ and not at $R_5$. The reason is that node 2 receives strong interference from node 4, which is near.

When the relays are clustered at the
destination, we expect $R_2$ to constrain the overall rate.  This is shown in Fig.~\ref{fig:rate_power_2hop_4}.  The reception rate at node 2 is low as the signal from node 1 is severely attenuated due to the large $d_{12}$ and high interference from nodes 4 and 5,
which are close to node 2.

It is noted that when the overall rate is constrained by $R_2$, the
power allocations affecting it, which are $\alpha_1$ and
$\alpha_3$ should be set to zero.  Setting $\alpha_1=0$, we ensure
that all power from node 1 carries new information to node 2.
Setting $\alpha_3=0$, we maximize the amount of interference that
node 2 can cancel in its decoding.

\subsubsection{Omniscient Coding} \label{sec:mrc_omniscient}
In omniscient decode-forward, encoding is as follows.
\begin{itemize}
\item Node 4 sends $X_4 = \sqrt{P_4}U_4$.
\item Node 3 sends $X_3 = \sqrt{(1-\alpha_3)P_3}U_3 + \sqrt{\alpha_3P_3}U_4$.
\item Node 2 sends $X_2 = \sqrt{(1-\alpha_2 - \beta_2)P_2}U_2 + \sqrt{
\beta_2P_2}U_3 + \sqrt{\alpha_2P_2}U_4$.
\item Node 1 sends $X_1 = \sqrt{(1 - \alpha_1 - \beta_1 -
\gamma_1)P_1}U_1 + \sqrt{\gamma_1P_1}U_2 + \sqrt{\beta_1P_1}U_3 + \sqrt{\alpha_1P_1}U_4$.
\end{itemize}
Here, $U_i, i=1, 2, 3, 4$ are independent Gaussian random variables with unit variances, $0 \leq \alpha_1 + \beta_1 + \gamma_1 \leq 1$, $0 \leq \alpha_2 + \beta_2 \leq 1$, $0 \leq \alpha_3 \leq 1$, and $\alpha_i, \beta_j, \gamma_1 \geq 0, i=1, 2, 3, j=1, 2$.  To illustrate the power splits, let us consider node 1.,It allocates $\alpha_1$ of its total power to
transmit to node 5, $\beta_1$ of its power to node 4, $\gamma_1$ of
its power to node 3, and the remaining power to node 2.

Fixing some $\{ \alpha_1, \beta_1, \gamma_1, \alpha_2, \beta_2,
\alpha_3 \}$, the reception rate at node 2 is
\begin{subequations}
\begin{align}
R_2 & = I(X_1;Y_2|X_2X_3X_4)\\
& = \frac{1}{2} \log 2\pi e \left[ \kappa d_{12}^{-\eta}
(1-\alpha_1-\beta_1-\gamma_1) P_1 + N_2\right] - \frac{1}{2} \log 2\pi e N_2\\
& = \frac{1}{2} \log \left[ 1 + d_{12}^{-2}(1-\alpha_1-\beta_1-\gamma_1)P_1\right].
\end{align}
\end{subequations}
Here, we have substituted $\kappa =1$, $\eta=2$, and
$N_2=1$W.  The reception rates at nodes 3, 4, and 5 can be computed in a similar way. Omniscient decode-forward achieves rates up to
\begin{equation}\label{eq:rate_complete_view}
R_{\text{omnicient}} = \max_{\{ \alpha_1, \beta_1, \gamma_1, \alpha_2, \beta_2,
\alpha_3 \}} \min_{t \in \{2,3,4,5\}} R_t.
\end{equation}

We define the following efficiency term to benchmark the performance of $k$-hop myopic coding.
\begin{equation}
\rho_k = \frac{R_{k-\text{hop}}}{R_{\text{omniscient}}},
\end{equation}
where $k \in \{1, 2, \dotsc, T-1\}$. It is the ratio of the maximum achievable rate of a $k$-hop myopic coding strategy to that of the corresponding omniscient coding strategy.

\begin{figure}[t]
\begin{minipage}[b]{0.48\linewidth}
\centering
\includegraphics[width=\textwidth]{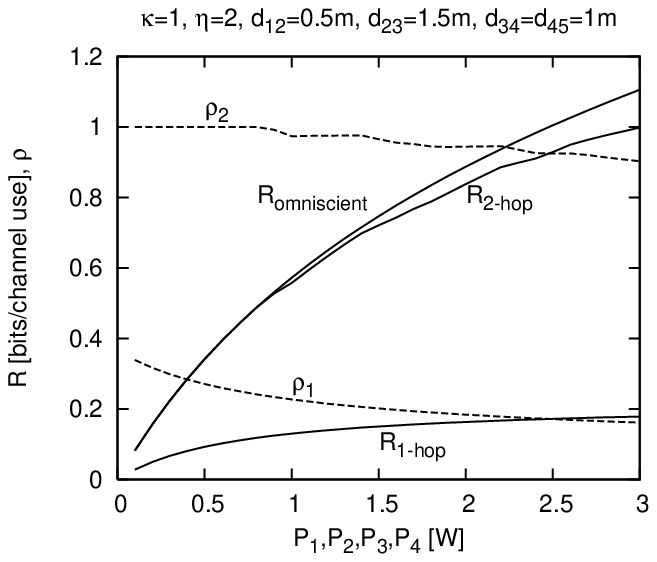}
\caption{Achievable rates under different coding strategies in the five-node
multiple-relay channel.} \label{fig:rate_power_5_node}
\end{minipage}
\hspace{0.3cm}
\begin{minipage}[b]{0.48\linewidth}
\centering
\includegraphics[width=\textwidth]{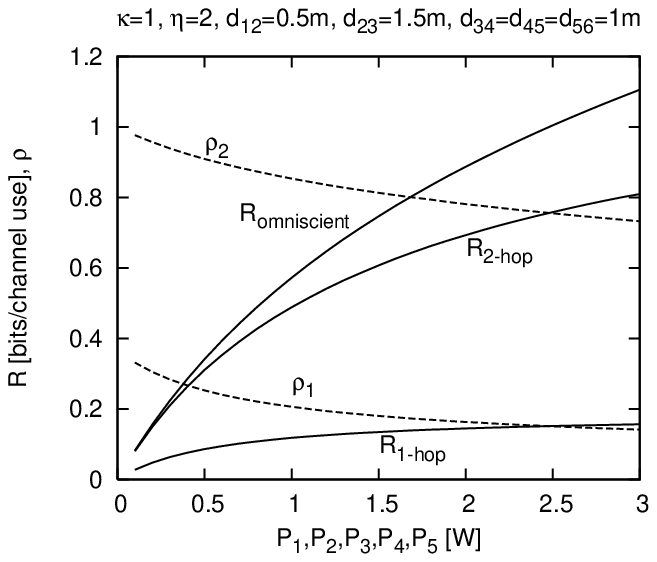}
\caption{Achievable rates under different coding strategies in the six-node
multiple-relay channel.} \label{fig:rate_power_6_node}
\end{minipage}
\end{figure}

Figs.~\ref{fig:rate_power_5_node} and \ref{fig:rate_power_6_node} show achievable rates in the five-node and the six-node multiple-relay channel respectively, using one-hop, two-hop, and omniscient decode-forward.

The maximum rate achievable by myopic coding can never exceed that by the corresponding omniscient coding. This is because under myopic coding, every node treats the transmissions of the nodes outside its view as noise. In addition, a node can only transmit limited messages. On the other hand, under omniscient coding, a node can decode the signals from all the nodes behind and cancel the transmissions of all the nodes in front. A node can also possibly transmit all previously decoded messages.

In Fig.~\ref{fig:rate_power_5_node}, we see a seemingly strange result that the maximum achievable rate of two-hop myopic decode-forward is as high as that of omniscient decode-forward. This can happen in a five-node channel under certain circumstances. Using either omniscient or two-hop myopic decode-forward, node 3 in the five-node multiple-relay channel can communicate with all other nodes, i.e., it decodes from nodes  1 and 2, and cancels transmissions from node 4. So, when the overall transmission rates is constrained by $R_3$, the maximum achievable rate of two-hop myopic decode-forward is the same as that of omniscient decode-forward. This explains why $\rho_2=1$ at low SNR in Fig.~\ref{fig:rate_power_5_node}.

However, as the number of relays increases, we expect achievable rates of two-hop myopic decode-forward to be strictly less than that of omniscient decode-forward. We see that this is indeed the case from Fig.~\ref{fig:rate_power_6_node}, in which $\rho_2$ is strictly less than 1.

Comparing achievable rates of one-hop and two-hop myopic decode-forward, the rates improve significantly when one more node is added into the nodes' view. This suggests that in a large network with many relays, $k$-hop myopic decode-forward, where $k$ needs not be large, could achieve rates close to that of omniscient decode-forward.

Furthermore, $\rho_1$ and $\rho_2$ are high in the low SNR regime. The efficiency drops as the SNR increases. To understand this phenomenon, we consider different types of noise, i.e., receiver noise and interference. The nodes in both omniscient and myopic decode-forward experience the same receiver noise. So, in the low SNR regime where the receiver noise is dominant, myopic decode-forward performs close to omniscient decode-forward, and the efficiency is higher. On the other hand, in the high SNR regime, the interference (which a node cannot cancel in myopic decode-forward but can in omniscient decode-forward) is dominant. So, the efficiency of myopic decode-forward drops.


\subsection{Extending to $k$-Hop Myopic Coding} \label{sec:k_hop}
Now, we generalize two-hop myopic decode-forward to $k$-hop myopic decode-forward where $k \in \{1,\dotsc,T-1\}$ and have the following theorem.

\begin{thm}\label{thm:k_hop}
Let
\begin{equation}
\Big( \mathcal{X}_1 \times \dotsm \times \mathcal{X}_{T-1}, p^*(y_2, \dotsc, y_T | x_1, \dotsc, x_{T-1}),
\mathcal{Y}_2 \times \dotsm \times \mathcal{Y}_T \Big)\nonumber
\end{equation}
be a $T$-node memoryless multiple-relay channel.  Under $k$-hop decode-forward, the rate $R$ is achievable, where
\begin{subequations} \label{eq:k-hop-myopic-rate}
\begin{align}
R &\leq \max_{\pi(\cdot)} \max_{p(\cdot)} \min_{t \in \{2, \dotsc, T\}}  I(U_{\pi(t-k)},\dotsc,U_{\pi(t-1)};Y_{\pi(t)}|U_{\pi(t)},\dotsc ,U_{\pi(t+k-1)})\\
&= R_{k\text{-hop}}.
\end{align}
\end{subequations}
Here, $U_{\pi(m)} = 0$, for all $m=2-k, 3-k, \dotsc, 0, T, T+1, \dotsc, T+k-1$. The outer maximization is over all relay permutations and the inner maximization is taken over all joint distributions of the form
\begin{subequations}
\begin{align}
& p(x_1, x_2 \dotsc, x_{T-1}, u_1, u_2 \dotsc, u_{T-1}, y_2, y_3 \dotsc, y_T)\nonumber\\
& = p(u_{\pi(1)})p(u_{\pi(2)})\dotsm p(u_{\pi(T-1)})\nonumber\\
& \quad \times p(x_{\pi(T-1)}|u_{\pi(T-1)})p(x_{\pi(T-2)}|u_{\pi(T-2)},u_{\pi(T-1)}) \dotsm  p(x_{\pi(T-k)}|u_{\pi(T-k)},u_{\pi(T-k+1)}\dotsc,u_{\pi(T-1)})\nonumber\\
& \quad \times p(x_{\pi(T-k-1)}|u_{\pi(T-k-1)},u_{\pi(T-k)} \dotsc, u_{\pi(T-2)}) \dotsm  p(x_{\pi(1)}|u_{\pi(1)},u_{\pi(2)},\dotsc,u_{\pi(k)})\nonumber\\
& \quad \times p^*(y_2, \dotsc, y_T | x_1, \dotsc, x_{T-1}). \nonumber
\end{align}
\end{subequations}
\end{thm}

The proof can be found in Appendix~\ref{append:k_hop}. In the extreme case where $k=T-1$, we end up with omniscient decode-forward.

\subsection{On the Gaussian Multiple Relay Channel with Fading}
In the analyses so far, we compared the performance of myopic coding strategies in static Gaussian channels, i.e., without fading. Now, we explain how myopic coding is done in the Gaussian channel with phase fading or Rayleigh fading.

It has been shown by Kramer \emph{et al.}~\cite[Theorem 8]{kramergastpar04} that under phase fading or Rayleigh fading, the maximum omniscient decode-forward rate can be achieved by independent Gaussian input distributions. In this case, $X_i, i=1, \dotsc, T-1$, are independent Gaussian random variables. Under omniscient decode-forward, node $t$ decodes from all nodes $i, i < j$, and cancels the transmissions of nodes $l, l \geq j$. In $k$-hop myopic decode-forward, the nodes transmit independent Gaussian signals as they would under the omniscient coding. However, in the decoding, node $t$ decodes the signals only from $k$ nodes behind, i.e., nodes $i, i= \max\{1, t-k\}, \dotsc, t-1$. It cancels the transmissions from only $k$ nodes in front (including itself), i.e., nodes $l, l=t, \dotsc, \min\{t+k-1,T-1\}$. It treats the rest of the transmissions as noise. The following theorem characterizes the performance of $k$-hop myopic decode-forward for the Gaussian multiple-relay channel with phase fading or Rayleigh fading.

\begin{thm}
Consider a $T$-node Gaussian multiple-relay channel with phase fading or Rayleigh fading.  Using $k$-hop decode-forward, the rate in equation \eqref{eq:k-hop-myopic-rate} is achievable, by setting $X_i = U_i, x_i = u_i, \forall i = 1, 2, \dotsc, T-1$. 
\end{thm}

The proof for the above theorem is straight forward given that the nodes transmit independent signals in the fading channel.

\subsection{Myopic Coding in Large Multiple-Relay Channels}\label{sec:t_node}
One potential problem of myopic coding is whether the rate vanishes when the number of nodes in the network grows.  This concern arises because in myopic decode-forward, a node treats transmissions of nodes beyond its view as pure noise.  As the number of transmitting nodes grows to infinity and each decoding node only has a limited view, the noise power might sum to infinity. The noise might overpower the signal power and drive the transmission rate to zero.

In this section, we scrutinize achievable rates of two-hop myopic decode-forward in the $T$-node multiple-relay channel when $T$ grows to infinity.  The rationale of studying two-hop myopic coding is that we can always achieve higher transmission rates using $k$-hop myopic coding with $k>2$.

\begin{thm}\label{thm:large_network}
Achievable rates of $k$-hop myopic decode-forward in the $T$-node Gaussian multiple-relay channel are bounded away from zero, for any $T \geq 3$.
\end{thm}

\begin{figure*}[t]
\centering
\includegraphics[width=16cm]{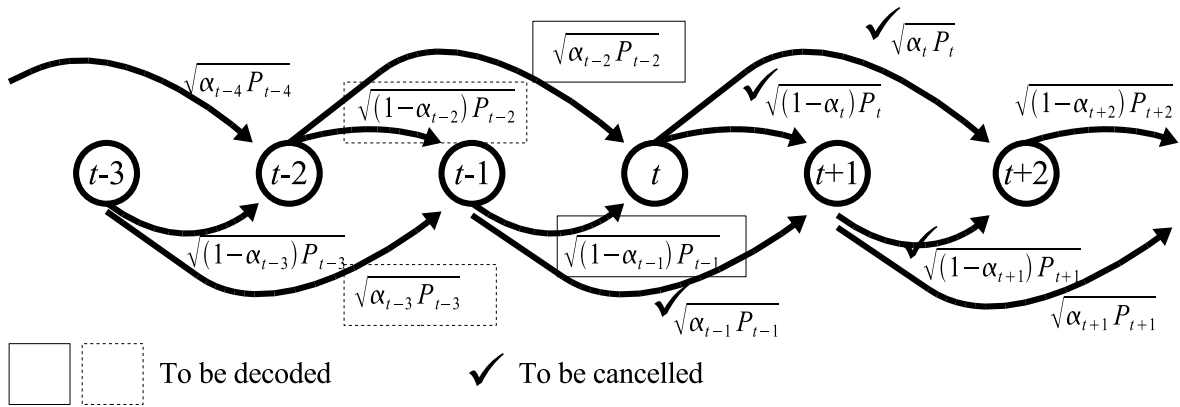}
\caption{The power allocation of two-hop myopic decode-forward for the Gaussian multiple-relay channel.} \label{fig:2_hop_gmrc}
\end{figure*}

Now, we prove Theorem~\ref{thm:large_network}. In two-hop myopic decode-forward for the $T$-node Gaussian multiple-relay channel (we shall extend $T$ to infinity later), the transmission of each node is as follows.
\begin{itemize}
\item Node $t, t=1, 2, \dotsc, T-2$, sends $X_t = \sqrt{\alpha_tP_t}U_{t+1} + \sqrt{(1-\alpha_t)P_t}U_t$.
\item Node $T-1$ sends $X_{T-1} = \sqrt{P_{T-1}}U_{T-1}$.
\end{itemize}
where $U_i, i=1, 2, \dotsc, T-1$, are independent Gaussian random variables with unit variances and $0 \leq \alpha_i \leq 1$.  The transmissions of the nodes around node $t$ are depicted in Fig.~\ref{fig:2_hop_gmrc}.

Assume that all the nodes are equally spaced at 1m apart and transmit at power $P$.  Consider the received signal power at node $t$, we can always find a non-empty set $\{ (\alpha_1, \dotsc, \alpha_{T-2}) : 0 \leq \alpha_i \leq 1, i=1, \dotsc, T-2 \}$ such that
\begin{subequations}
\begin{align}
P_{\text{sig}}(t) & = \left( \sqrt{ 3^{-\eta}\alpha_{t-3} \kappa P}+ \sqrt{2^{-\eta}(1-\alpha_{t-2})\kappa P} \right)^2   + \left( \sqrt{ 2^{-\eta}\alpha_{t-2}\kappa P} + \sqrt{ 1^{-\eta}(1-\alpha_{t-1}) \kappa P}\right)^2\\
& = \left( \sqrt{ 3^{-\eta}\alpha_{t-3}\kappa P} + \sqrt{2^{-\eta}(1-\alpha_{t-2})\kappa P} \right)^2   + \left( \sqrt{ 2^{-\eta}\alpha_{t-2}\kappa P} + \sqrt{ 1^{-\eta}(1-\alpha_{t-1})\kappa P}\right)^2\\
& > 0,
\end{align}
\end{subequations}
for $t \geq 4$, and
\begin{subequations}
\begin{align}
P_{\text{sig}}(2) & = (1-\alpha_{1})\kappa P > 0\\
P_{\text{sig}}(3) & = 2^{-\eta}(1-\alpha_{1})\kappa P + \left( \sqrt{ 2^{-\eta}\alpha_{1}\kappa P} + \sqrt{ 1^{-\eta}(1-\alpha_{2})\kappa P}\right)^2 > 0.
\end{align}
\end{subequations}

Now we consider nodes $4 \leq t \leq T-3$, the noise power is $P_{\text{noise}}(t) = N_t < \infty$, and the interference power is given by
\begin{subequations}
\begin{align}
P_{\text{int}}(t)& = \left( \sqrt{ 3^{-\eta}(1-\alpha_{t-3})\kappa P} + \sqrt{4^{-\eta}\alpha_{t-4}\kappa P} \right)^2  + \left( \sqrt{ 4^{-\eta}(1-\alpha_{t-4})\kappa P} + \sqrt{5^{-\eta}\alpha_{t-5}\kappa P} \right)^2 + \dotsm \nonumber\\
& \quad + \left( \sqrt{ (t-2)^{-\eta}(1-\alpha_{2})\kappa P} + \sqrt{(t-1)^{-\eta}\alpha_{1}\kappa P} \right)^2  + (t-1)^{-\eta}(1-\alpha_{1})\kappa P \nonumber\\
& \quad + \left( \sqrt{ 1^{-\eta}\alpha_{t+1}\kappa P} + \sqrt{2^{-\eta}(1-\alpha_{t+2})\kappa P} \right)^2  + \left( \sqrt{ 2^{-\eta}\alpha_{t+2}\kappa P} + \sqrt{3^{-\eta}(1-\alpha_{t+3})\kappa P} \right)^2 + \dotsm \nonumber\\
& \quad + \Big( \sqrt{ (T-t-3)^{-\eta}\alpha_{T-3}\kappa P}  + \sqrt{(T-t-2)^{-\eta}(1-\alpha_{T-2})\kappa P} \Big)^2\nonumber\\
& \quad + \left( \sqrt{ (T-t-2)^{-\eta}\alpha_{T-2}\kappa P} + \sqrt{(T-t-1)^{-\eta}\kappa P} \right)^2,
\end{align}
\end{subequations}

\begin{subequations}
\begin{align}
\frac{P_{\text{int}}(t)}{\kappa P} & = 3^{-\eta}\alpha_{t-3} + 4^{-\eta} + 5^{-\eta} + \dotsm + (t-1)^{-\eta} \nonumber\\
& \quad  + 2\sqrt{ 3^{-\eta}4^{-\eta}(1-\alpha_{t-3})\alpha_{t-4} } + 2\sqrt{ 4^{-\eta}5^{-\eta}(1-\alpha_{t-4})\alpha_{t-5} }  + \dotsm  + 2\sqrt{ (t-2)^{-\eta}(t-1)^{-\eta}(1-\alpha_2)\alpha_1 } \nonumber\\
& \quad + 1^{-\eta}\alpha_{t+1} + 2^{-\eta} + 3^{-\eta} + \dotsm + (T-t-1)^{-\eta} \nonumber\\
& \quad  + 2\sqrt{ 1^{-\eta}2^{-\eta}\alpha_{t+1}(1-\alpha_{t+2}) }  + 2\sqrt{ 2^{-\eta}3^{-\eta}\alpha_{t+2}(1-\alpha_{t+3}) }  + \dotsm \nonumber\\
& \quad  + 2\sqrt{ (T-t-3)^{-\eta}(T-t-2)^{-\eta}\alpha_{T-3}(1-\alpha_{T-2}) }.
\end{align}
\end{subequations}

Simplifying, we get
\begin{subequations}
\begin{align}
\frac{P_{\text{int}}(t)}{\kappa P} & = 3^{-\eta}\alpha_{t-3} + \sum_{j=4}^{t-1} \frac{1}{j^{\eta}} + 1^{-\eta}\alpha_{t+1} + \sum_{j=2}^{T-t-1} \frac{1}{j^{\eta}}   + 2 \sum_{j=3}^{t-2} \sqrt{ \frac{(1-\alpha_{t-j})\alpha_{t-(j+1)}}{j^{\eta}(j+1)^{\eta}} } + 2 \sum_{j=1}^{T-t-3} \sqrt{ \frac{\alpha_{t+j}(1-\alpha_{t+j+1})} {j^{\eta}(j+1)^{\eta}} }\\
& < \sum_{j=3}^{t-1} \frac{1}{j^{\eta}} + \sum_{j=1}^{T-t-1} \frac{1}{j^{\eta}} + 2\sum_{j=3}^{t-2} \frac{1}{j^{\eta}} + 2\sum_{j=1}^{T-t-3} \frac{1}{j^{\eta}}\\
& < 6 \sum_{j=1}^{T} \frac{1}{j^{\eta}}< 6 \zeta(\eta).
\end{align}
\end{subequations}
Here $\zeta(\eta) = \sum_{j=1}^\infty \frac{1}{j^\eta}$ is the Riemann zeta function. It has been calculated that $\zeta(2)=\frac{\pi^2}{6}$, $\zeta(3)=1.202057...$ etc. It is easily seen that the Riemann zeta function is a decreasing function of $\eta$. Since, $\eta \geq 2$, $P_{\text{int}}(t) < \pi^2\kappa P$ for $4 \leq t \leq T-3$. We can also show that $P_{\text{int}}(t)/(\kappa P)$ for $t=2,3, T-2, T-1, T$ are bounded. Hence, we can always find a non-empty set $\{(\alpha_1, \dotsc, \alpha_{T-2})\}$ such that the reception rate at every node $t$, $\forall t \in \{2, 3, \dotsc, T \}$, is
\begin{equation}
R_t = \frac{1}{2} \log \left[ 1 + \frac{P_{\text{sig}}(t)}{P_{\text{int}}(t) + N_t} \right] > 0,
\end{equation}
which is bounded away from zero. This means the maximum achievable rate
\begin{equation}
R_{\text{2-hop}} = \max_{\{\alpha_1, \dotsc, \alpha_{T-2}\}} \min_{t \in \{2, 3, \dotsc, T \}} R_t > 0
\end{equation}
is bounded away from zero.

When more nodes are included in the view of myopic coding, $P_{\text{sig}}$ increases and $P_{\text{int}}$ decreases. In general, assuming that the nodes are roughly equally spaced, achievable rates of myopic decode-forward  are bounded away from zero even when the network size grows to infinity.

In the next two sections, we study achievable rates of myopic and omniscient coding strategies for the multiple-access relay channel and the broadcast relay channel.

\section{Myopic Coding in the Multiple-Access Relay Channel}\label{sec:myopic-marc}
\subsection{Channel Model}
The multiple-access relay channel has multiple sources, one relay, and one destination. In the $T$-node multiple-access relay channel, nodes 1 to $T-2$ are the sources, node $T-1$ is the relay, and node $T$ is the destination. The rates $(R_1, \dotsc, R_{T-2})$ for nodes $1, \dotsc, T-2$ respectively are said to be achievable if each node can transmit messages to the destination at their respective rates with diminishing error probability. They follow closely the definition that we adopt for the multiple-relay channel. The sources do not receive feedback from the channel. 
The multiple-access relay channel can be completely described by its channel distribution of the following form.
\begin{equation}
p^*(y_{T-1},y_T | x_1, \dotsc, x_{T-1}).
\end{equation}

\subsection{Achievable Rates}
In this paper, we consider the four-node multiple-access relay channel, where nodes 1 and 2 are the sources, node 3 is the relay, and node 4 is the destination.  We assume that data from node 1 and node 2 are independent.
We investigate decode-forward based coding strategies for the multiple-access relay channel, in which the relay must decode all messages from both sources.

\begin{figure}[t]
\begin{minipage}[b]{0.48\linewidth}
\centering
\includegraphics[width=4cm]{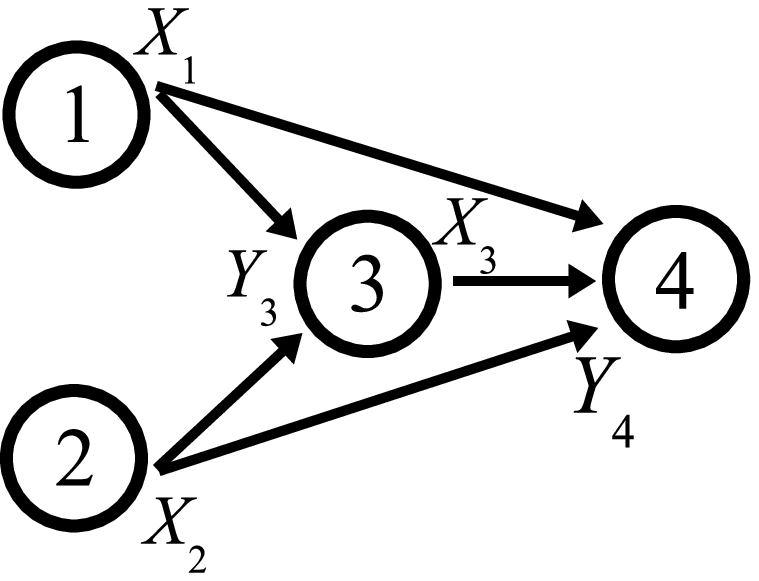}
\caption{Omniscient decode-forward for the four-node multiple-access relay channel.} \label{fig:4SensorLong}
\end{minipage}
\hspace{0.3cm}
\begin{minipage}[b]{0.48\linewidth}
\centering
\includegraphics[width=4cm]{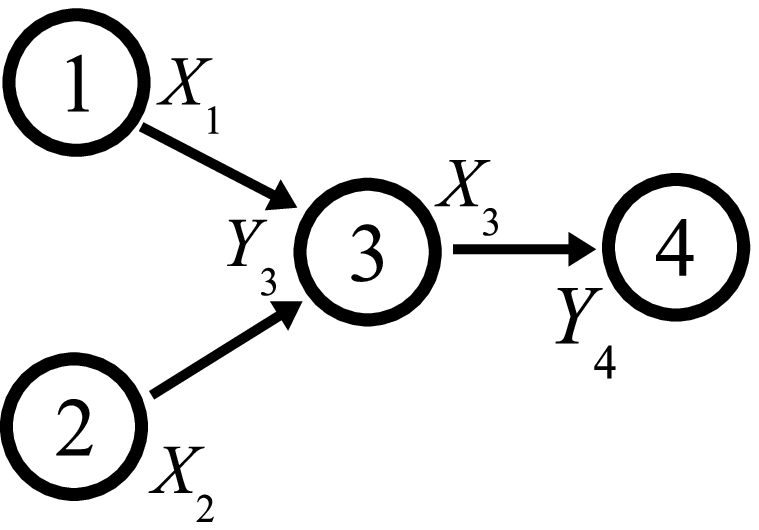}
\caption{One-hop myopic decode-forward for the four-node multiple-access relay channel.} \label{fig:4SensorShort}
\end{minipage}
\end{figure}

\subsubsection{Omniscient Coding}
In omniscient decode-forward for the four-node multiple-access relay channel, nodes 1 and
2 transmit to both nodes 3 and 4.  This is depicted in
Fig.~\ref{fig:4SensorLong}. Using offset encoding \cite{sankara04b} and sliding window decoding, omniscient decode-forward achieves the following rate region~\cite{kramergastpar04}.
\begin{subequations}\label{eq:ratesLong}
\begin{align}
R_1 & \leq I(X_1;Y_3|U_1,U_2,X_2,X_3)\\
R_1 & \leq I(X_1,X_3;Y_4|U_2,X_2)\\
R_2 & \leq I(X_2;Y_3|U_1,U_2,X_1,X_3)\\
R_2 & \leq I(X_2,X_3,Y_4|U_1,X_1)\\
R_1 + R_2 & \leq I(X_1,X_2;Y_3|U_1,U_2,X_3)\label{eq:sumRatesLong1}\\
R_1 + R_2 & \leq I(X_1,X_2,X_3;Y_4),\label{eq:sumRatesLong2}
\end{align}
\end{subequations}
where the mutual information terms are taken over
\begin{equation}
p(u_1, u_2, x_1,x_2, x_3, y_3, y_4) = p(u_1, x_1) p(u_2, x_2)p(x_3|u_1, u_2)
p^*(y_3,y_4|x_1,x_2,x_3).
\end{equation}
We note that in this four-node multiple-access relay channel, two-hop myopic decode-forward is equivalent to  omniscient decode-forward.

\subsubsection{One-Hop Myopic Coding}
In one-hop myopic decode-forward for the four-node multiple-access relay channel, nodes 1 and 2 transmit to node 3, but not to node 4. In this scenario, we have the channel model as
depicted in Fig.~\ref{fig:4SensorShort}. We can view this as a multiple-access channel (from nodes 1--2 to node 3) cascaded with a point-to-point channel (from node 3 to node 4). Modifying the results of the
multiple-access channel in \cite{verdu98}, the following rate region is achievable by one-hop myopic decode-forward.
\begin{subequations}\label{eq:ratesShort}
\begin{align}
R_1 & \leq I(X_1;Y_3|X_2,X_3)\\
R_2 & \leq I(X_2;Y_3|X_1,X_3)\\
R_1 + R_2 & \leq I(X_1,X_2;Y_3|X_3) \label{eq:sumRatesShort1}\\
R_1 + R_2 & \leq I(X_3;Y_4), \label{eq:sumRatesShort2}
\end{align}
\end{subequations}
where the mutual information terms are derived under the joint
distributions $p(x_1,x_2, x_3,y_3,y_4) = p(x_1)p(x_2)p(x_3)$\\$p^*(y_3,y_4|x_1,x_2,x_3)$.

\subsection{Performance Comparison}
\subsubsection{Channel Setup}

Now, we investigate achievable rates of one-hop myopic decode-forward and omniscient decode-forward for the four-node Gaussian multiple-access relay channel. Nodes 1, 2, and 3 send $X_1$, $X_2$, and $X_3$ respectively.  Node 3 receives
\begin{equation}\label{eq:powerAt3}
Y_3 = \sqrt{\kappa d_{13}^{-\eta}}X_1 + \sqrt{\kappa d_{23}^{-\eta}}X_2 + Z_3
\end{equation}
and node 4 receives
\begin{equation}\label{eq:powerAt4}
Y_4 = \sqrt{\kappa d_{14}^{-\eta}}X_1 + \sqrt{\kappa d_{24}^{-\eta}}X_2 +
\sqrt{\kappa d_{34}^{-\eta}}X_3 + Z_4
\end{equation}
where $Z_3$ and $Z_4$ are independent zero-mean white Gaussian noise
with variances $N_3$ and $N_4$ respectively. $X_1$, $X_2$, and
$X_3$ are zero-mean Gaussian random variables with fixed average transmit power $E[X_i^2] = P_i, \text{ }i=1,2,3$.
In our analysis, we use the following parameters.
$d_{12} = d_{23} = d_{13} = 1$m, $N_3 = N_4 = 1$W, $\kappa=1$, $\eta
= 2$, $d_{13}=d_{23}$, and $d_{14}=d_{24}$. We let
$R_3'$ be the reception rate (sum rate) at node 3,
and $R_4'$ the reception rate (sum rate) at node
4.

\subsubsection{One-Hop Myopic Coding}

From \eqref{eq:sumRatesShort1}, the reception rate (sum rate) at node 3 is
\begin{subequations}
\begin{align}
R'_3 & = \frac{1}{2}\log 2\pi e E[Y^2_3] - \frac{1}{2}\log 2\pi e
E[Z_3^2]\\
& = \frac{1}{2} \log 2\pi e \Bigl( \kappa d_{13}^{-\eta} P_1 +
\kappa d_{23}^{-\eta} P_2 + N_3 \Bigr) - \frac{1}{2} \log 2\pi e N_3\\
& = \frac{1}{2} \log ( 1 + P_1 + P_2 ). \label{eq:substitution1}
\end{align}
\end{subequations}
Here, we have substituted $\kappa=1$, $d_{13}=d_{23} = 1$m, $\eta=2$, and $N_3=1$W.
From \eqref{eq:sumRatesShort2}, the reception rate at node 4 is
\begin{subequations}
\begin{align}
R'_4 & = \frac{1}{2} \log 2\pi e \Bigl( \kappa d_{14}^{-\eta}P_1 + \kappa d_{24}^{-\eta}P_2
+ \kappa d_{34}^{-\eta}P_3 + N_4 \Bigr)  - \frac{1}{2} \log 2\pi e \Bigl( \kappa d_{14}^{-\eta}P_1+ \kappa d_{24}^{-\eta}
P_2 + N_4 \Bigr)\\
& = \frac{1}{2} \log \left( 1+ \frac{P_3/d_{34}^2}{1+P_1/d_{14}^2 +
P_2/d_{24}^2} \right)\label{eq:substitution2}
\end{align}
\end{subequations}
where \eqref{eq:substitution2} is obtained after
substituting $\kappa=1$, $\eta=2$, $N_4=1$W, and $d_{14}^2 = d_{24}^2
= \left( \frac{\sqrt{3}}{2} + d_{34} \right)^2 + \frac{1}{4}$.

Since each message must be completely decoded by nodes 3 and 4, the following rates are achievable
\begin{equation}
R'=R_1+R_2 \leq \min \{ R'_3, R'_4\} = R_\text{1-hop}.
\end{equation}

\begin{figure}[t]
\begin{minipage}[b]{0.48\linewidth}
\centering
\includegraphics[width=\textwidth]{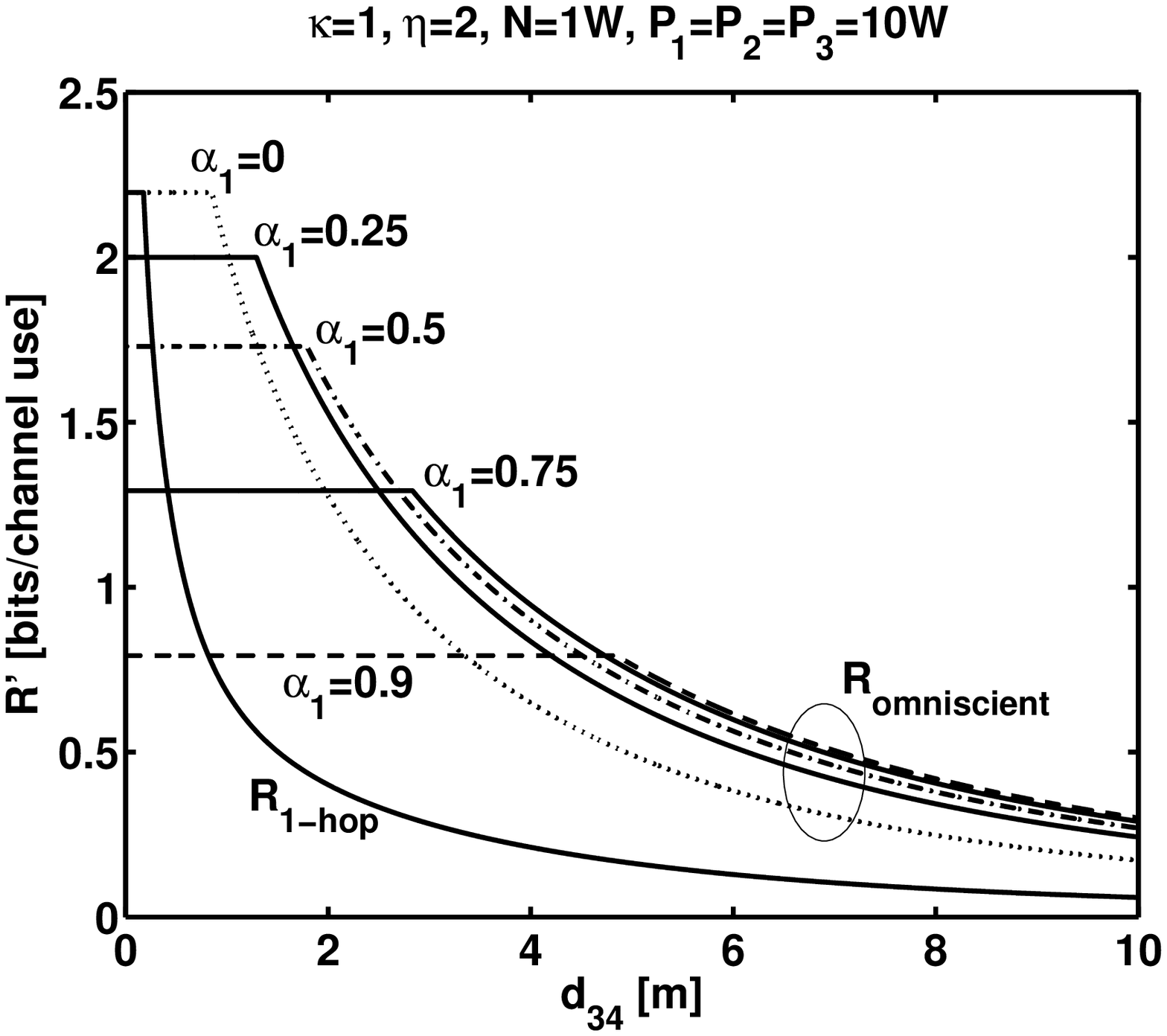}
\caption{Achievable sum rates of one-hop myopic decode-forward and omniscient decode-forward for the four-node multiple-access relay channel.} \label{fig:SumRateDistance_2}
\end{minipage}
\hspace{0.3cm}
\begin{minipage}[b]{0.48\linewidth}
\centering
\includegraphics[width=\textwidth]{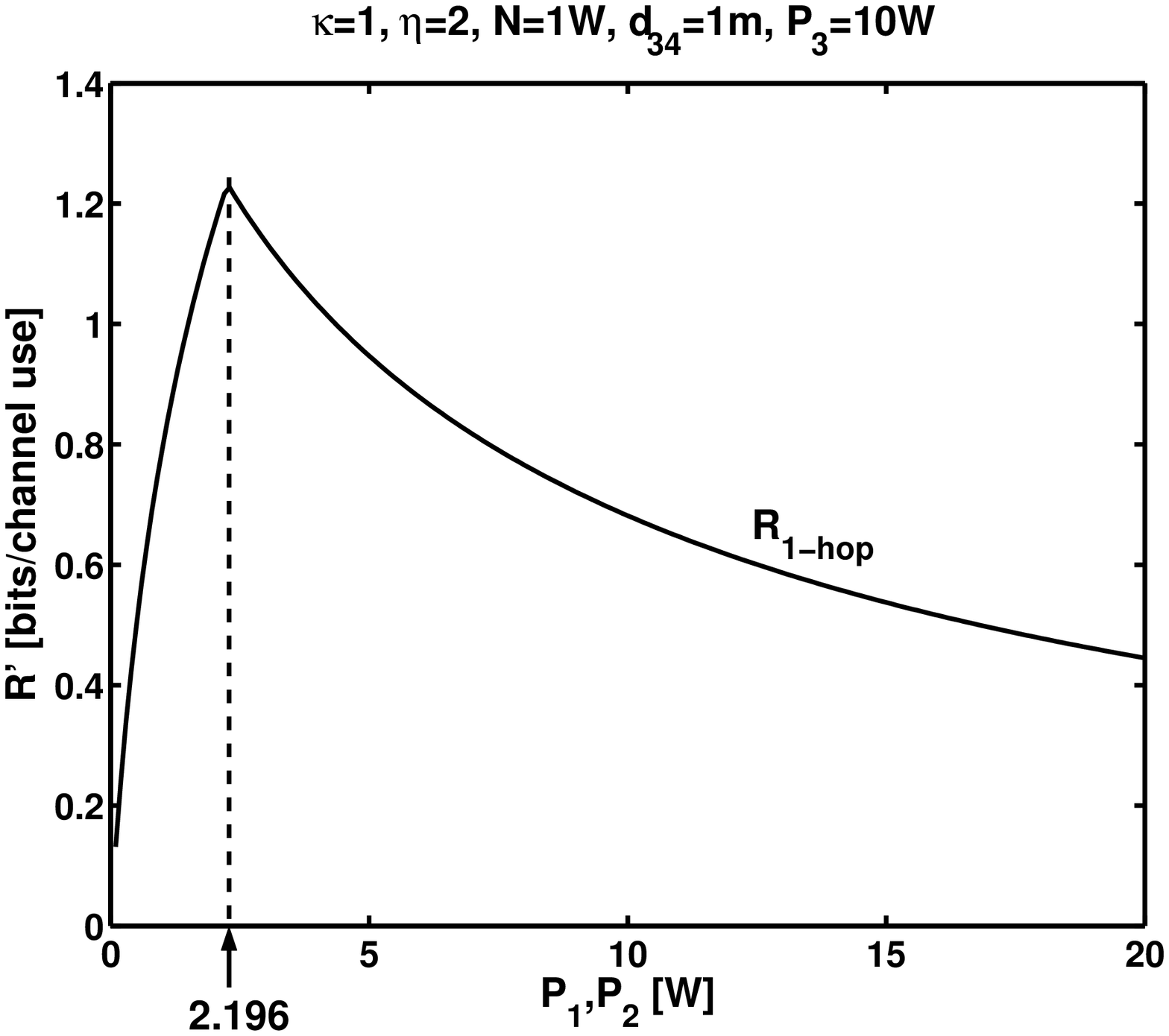}
\caption{Achievable sum rates of one-hop myopic decode-forward for the four-node multiple-access relay channel.} \label{fig:SumRatePower12}
\end{minipage}
\end{figure}

Fig.~\ref{fig:SumRateDistance_2} shows how the maximum achievable sum rate
$R_\text{1-hop}$ varies with $d_{34}$ when $P_1=P_2=P_3=10$W. When the
destination is near the relay, $R_4'$ is
higher than $R_3'$, which is a constant at
$I(X_1X_2;Y_3|X_3)=2.196$ bits/channel use. Hence, $R_\text{1-hop}$ is constrained by $R'_3$. When $d_{34}$ increases, $R_\text{1-hop}$ is
constrained by $R'_4$, which decreases as
$d_{34}$ increases.

Intuitively, when the rate is constrained by $R'_4$,
nodes 1 and 2 can reduce their transmit power to reduce the
interference from nodes 1 and 2 at node 4.
Fig.~\ref{fig:SumRatePower12} shows achievable rates when we vary  $P_1=P_2$ while keeping $d_{34}$ and
$P_3$ constant. When $P_1=P_2 \leq 2.196$W, $R_\text{1-hop}$ is constrained by $R'_3$. Increasing $P_1$ and $P_2$ increases $R_\text{1-hop}$. However, when $P_1$ and $P_2$ are large,
the interference at node 4 increases and $R_\text{1-hop}$ is now
constrained by $R'_4$. In this case, increasing $P_1$ and
$P_2$ decreases $R_\text{1-hop}$. We see that there is an
optimal point $P_1 = P_2 = 2.196$W for which $R_\text{1-hop}$ is maximized for fixed $d_{34}$ and $P_3$.

\subsubsection{Omniscient Coding}
In omniscient decode-forward, nodes 1, 2 and 3 transmit the following~\cite{kramer00}.
\begin{subequations}
\begin{align}
X_1 & = \sqrt{P_1} ( \sqrt{\alpha_1} U_1 + \sqrt{1-\alpha_1} V_1)\\
X_2 & = \sqrt{P_2} ( \sqrt{\alpha_2} U_2 + \sqrt{1-\alpha_2} V_2)\\
X_3 & = \sqrt{P_3} ( \sqrt{\beta_1} U_1 + \sqrt{\beta_2} U_2)
\end{align}
\end{subequations}
where $U_k$ and $V_k$, $k=1,2$, are independent, zero-mean Gaussian
random variables with unit variance, $0 \leq \alpha_1, \alpha_2 \leq
1$, $\beta_1, \beta_2 \geq 0$, and $\beta_1 + \beta_2 = 1$.

From \eqref{eq:sumRatesLong1}, the reception rate (sum rate) at node 3
is
\begin{subequations}
\begin{align}
 R'_3 & = H(Y_3|U_1,U_2,X_3) - H(Y_3|U_1,U_2,X_1,X_2,X_3)\\
& = \frac{1}{2} \log 2\pi e \bigl[ P_1\kappa d_{13}^{-\eta}(1-\alpha_1) +
P_2\kappa d_{23}^{-\eta}(1-\alpha_2) + N_3 \bigr] - \frac{1}{2} \log 2\pi e N_3\\
& = \frac{1}{2} \log \bigl[ 1+ P_1(1-\alpha_1) + P_2(1-\alpha_2)
\bigr].\label{eq:substitution3}
\end{align}
\end{subequations}
Here, \eqref{eq:substitution3} is obtained by substituting
$\kappa =1$, $d_{13} = d_{23} = 1$m, $N_3 = 1$W.

From \eqref{eq:sumRatesLong2}, the reception rate at node 4
is
\begin{subequations}
\begin{align}
R'_4 & = H(Y_4) - H(Y_4|X_1,X_2,X_3)\\
& = \frac{1}{2} \log 2\pi e \Bigg[ \frac{P_1}{d_{14}^2} +
\frac{P_2}{d_{24}^2} + \frac{P_3}{d_{34}^2}  + 2k\sqrt{P_1P_3(d_{14}d_{34})^{-\eta}\alpha_1\beta_1}   + 2k\sqrt{P_2P_3(d_{24}d_{34})^{-\eta}\alpha_2\beta_2}  + N_4
\Bigg] \nonumber\\
& \quad + \frac{1}{2} \log 2\pi e N_4\\
& = \frac{1}{2} \log \Bigg[ 1 + \frac{P_1}{d_{14}^2} +
\frac{P_2}{d_{24}^2} + \frac{P_3}{d_{34}^2} +
\frac{2\sqrt{\alpha_1\beta_1P_1P_3}}{d_{14}d_{34}} + \frac{2\sqrt{\alpha_2\beta_2P_2P_3}}{d_{24}d_{34}} \Bigg].
\end{align}
\end{subequations}
Here, we have substituted $\kappa =1$, $\eta = 2$, $N_4 = 1$W.
$d_{14}^2 = d_{24}^2 = \left( \frac{\sqrt{3}}{2} + d_{34} \right)^2
+ \frac{1}{4}$.

The following rates are achievable
\begin{equation}
R' = R_1+R_2 \leq \min \{ R'_3, R'_4\} = R_\text{omniscient} = R_\text{2-hop},
\end{equation}
for some $0 \leq \alpha_1, \alpha_2 \leq
1$ and $\beta_1 + \beta_2 = 1$.

To compare achievable rates of one-hop myopic decode-forward with that of omniscient decode-forward, we have calculated $R'$ for $P_1 = P_2 = P_3 =
10$W. Because of symmetry, we set $\alpha_1 = \alpha_2$ and $\beta_1 =
\beta_2 = \frac{1}{2}$.

Fig.~\ref{fig:SumRateDistance_2} shows achievable rates for
varying $d_{34}$ and $\alpha_1$ ($=\alpha_2$).  We see that when $d_{34}$ is small, i.e., the destination is close to the relay,
the optimal $\alpha_1$ is 0. This is intuitive because as $d_{34}$
is small, the overall rate is constrained by $R'_3$.  The
relay-to-destination link is almost noise free. The reception rate at node 3, $R'_3$, is maximized at $\alpha_1 = 0$ when nodes 1 and 2
allocate all signal power for new information (rather than helping the relay to transmit old information).

When $d_{34}$ is small, the maximum achievable sum rate of one-hop myopic decode-forward is the same as that of omniscient decode-forward. As the constraint is on $R'_3$, whether node 4 decodes additional signals from nodes 1 and 2 does not have any effect on the overall achievable rate. However, as $d_{34}$
increases, the rate constraint shifts to $R'_4$.
$R'_4$ of one-hop myopic decode-forward is lower than that of omniscient decode-forward because node 4 does not decode transmissions from nodes 1 and 2 in the former.

Also, when the maximum achievable sum rate is constrained by $R'_4$, the rate can be increased with a larger $\alpha_1$.  This is because
$\alpha_1$ controls the portion of power for direct transmission
from nodes 1 and 2 to node 4. Using a higher $\alpha_1$, the rate on
the constrained link $(1,2,3) \rightarrow 4$ improves and so does the overall rate. When the relay is close to the
destination, a smaller $\alpha_1$ is preferred. When the relay is
far away from the destination, higher achievable rates are possible using a larger $\alpha_1$. We note that no matter how far the relay is from the destination, the optimal $\alpha_1$ is always strictly less than 1. Setting $\alpha_1=1$ means the source does not send new information and merely repeats what the relay sends and hence new information is never transmitted.

\begin{figure}[t]
\begin{minipage}[b]{0.48\linewidth}
\centering
\includegraphics[width=\textwidth]{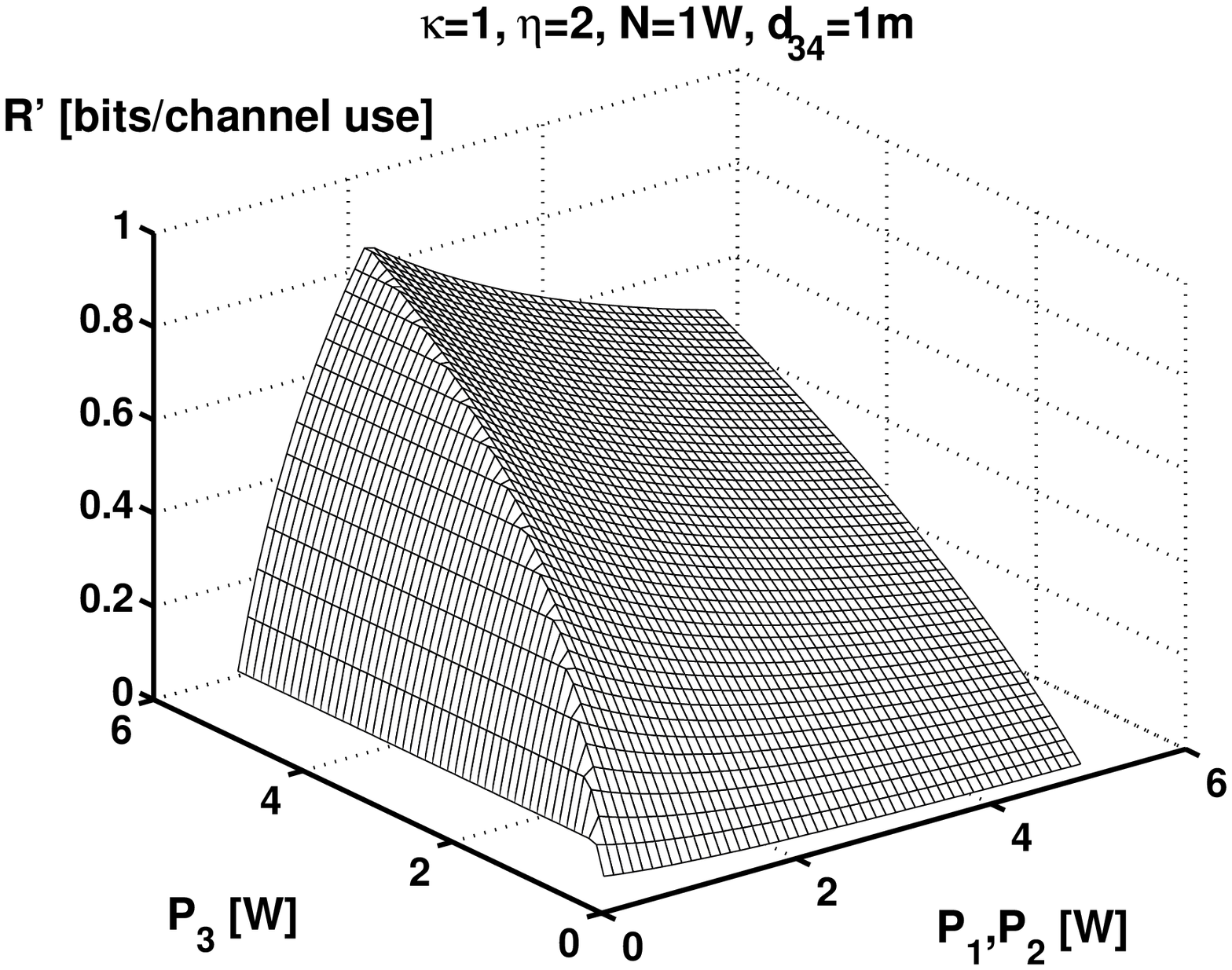}
\caption{$R'$ vs. $P_1,P_2$ and $P_3$ for one-hop myopic decode-forward for the four-node multiple-access relay channel.} \label{fig:SumRateSch1}
\end{minipage}
\hspace{0.3cm}
\begin{minipage}[b]{0.48\linewidth}
\centering
\includegraphics[width=\textwidth]{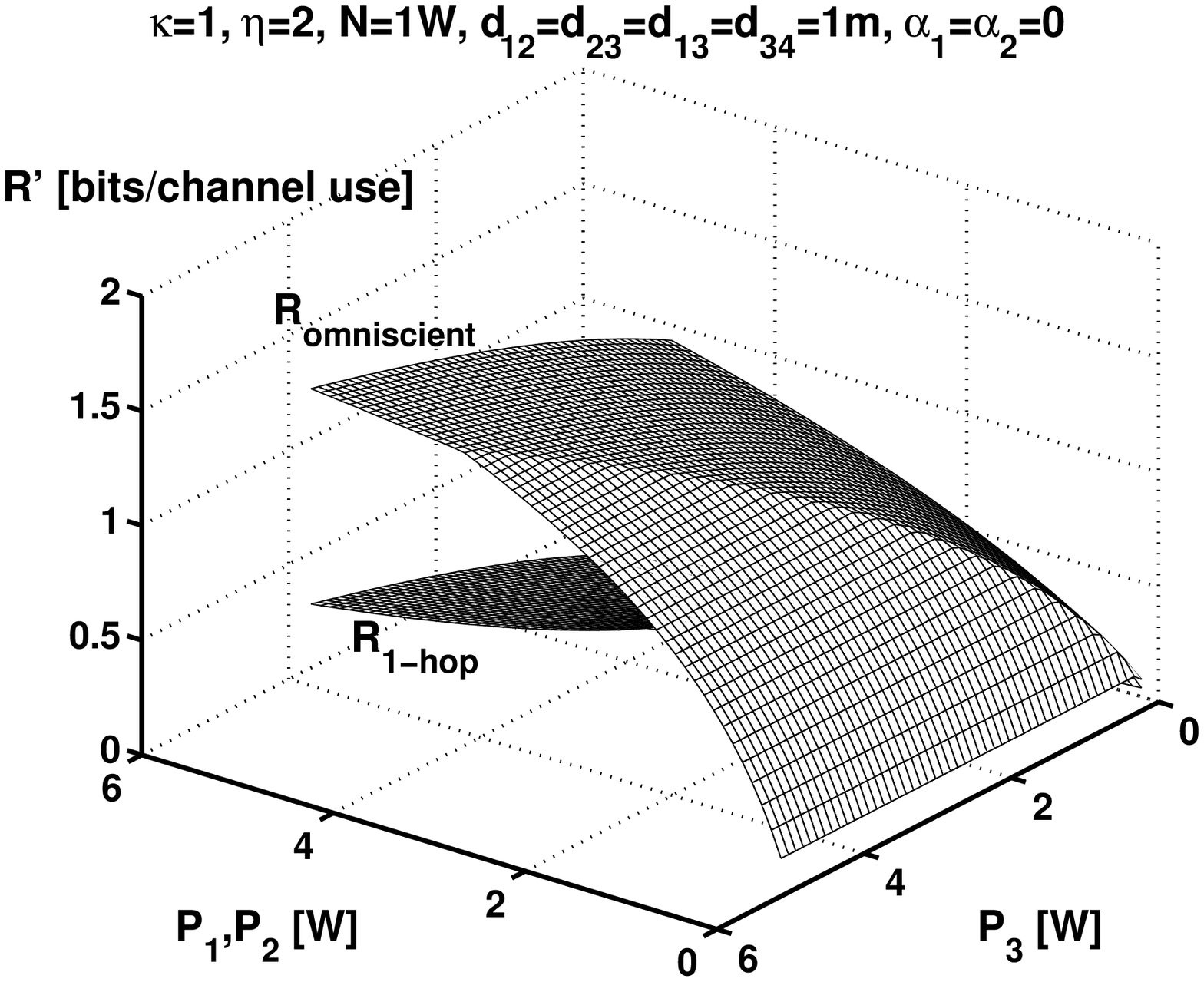}
\caption{Comparison of achievable sum rates of one-hop myopic decode-forward and omniscient decode-forward for the four-node multiple-access relay channel.} \label{fig:SumRateSch1&2}
\end{minipage}
\end{figure}

Figures~\ref{fig:SumRateSch1} and \ref{fig:SumRateSch1&2} depict
achievable sum rates of one-hop myopic decode-forward and omniscient decode-forward (with $\alpha_1 = \alpha_2
=0$ in the omniscient coding) for different transmission power.
$d_{34}$ is set to 1m. It is noted that for small $d_{34}$, the
optimal $\alpha_1$ and $\alpha_2$ are 0. So, we set  $\alpha_1=\alpha_2=0$ for the omniscient coding strategy.

In Fig.~\ref{fig:SumRateSch1}, we see that increasing $P_3$ always
increases achievable rates of both myopic decode-forward and omniscient decode-forward. This is because transmissions from node 3 are never treated as noise.  However, in one-hop myopic decode-forward, increasing
$P_1$ and $P_2$ decreases $R'_4$ and $R'$, as node 4 treats these transmissions as noise.  On the other hand, increasing the transmit power at any node always increases achievable rates in omniscient decode-forward, as all transmissions are either canceled off or decoded.

From Fig.~\ref{fig:SumRateSch1&2}, we see that when the sources transmit at low power and the relay transmits at high power, achievable sum rates of one-hop myopic decode-forward are as high as that of omniscient decode-forward. The reason for this is similar to that explain in Section~\ref{sec:mrc_omniscient}. When the source-relay link is the bottleneck of the overall transmission, achievable rates of myopic decode-forward are the same as that of the corresponding omniscient decode-forward.

\section{Myopic Coding in the Broadcast Relay Channel}\label{sec:myopic-brc}
\subsection{Channel Model}
The broadcast relay channel has one source, one relay, and multiple destinations. In a $T$-node broadcast relay channel, nodes 1 is the source (which does not receive feedback from the channel), node $2$ the relay, and nodes $3--T$ the destinations. The common rate $R_0$ (information that is common to all destinations) and the private rates $(R_3, \dotsc, R_{T})$ for nodes $3, \dotsc, T$ respectively are said to be achievable if the source can transmit information to the destinations at these rates with diminishing error probability.

The broadcast relay channel can be completely described by its channel distribution of the following form.
\begin{equation}
p^*(y_2, \dotsc, y_T | x_1, x_2).
\end{equation}

\subsection{Achievable Rates}
In this paper, we consider the four-node broadcast relay channel, where nodes 1 is the source, node 2 is the relay, and nodes 3 and 4 are the destinations. Node 1 is connected to a message generator that generates messages $W_3$ and $W_4$ to be sent to nodes 3 and 4 respectively; and common message $W_0$ to be sent to both destinations. We assume that $W_3$ and $W_4$ are independent.
Again, we use decode-forward-based coding strategies, in which the relay fully decodes all messages from the source.

\begin{figure}[t]
\begin{minipage}[b]{0.48\linewidth}
\centering
\includegraphics[width=4cm]{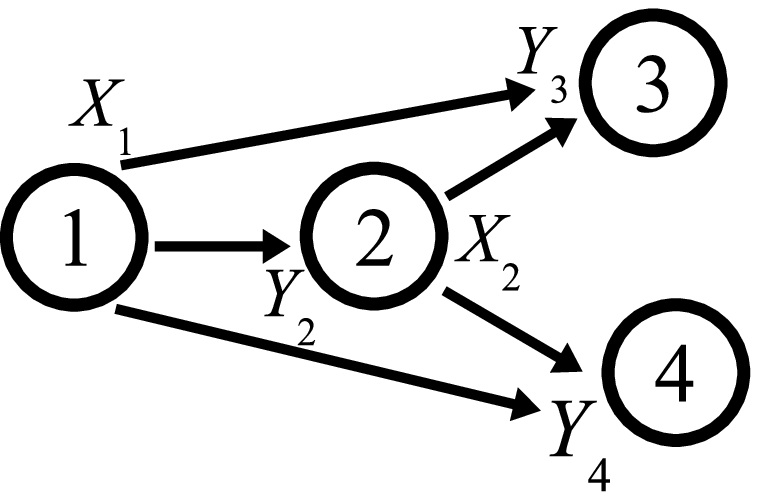}
\caption{Omniscient decode-forward for the four-node broadcast relay channel.} \label{fig:brc-omniscient}
\end{minipage}
\hspace{0.3cm}
\begin{minipage}[b]{0.48\linewidth}
\centering
\includegraphics[width=4cm]{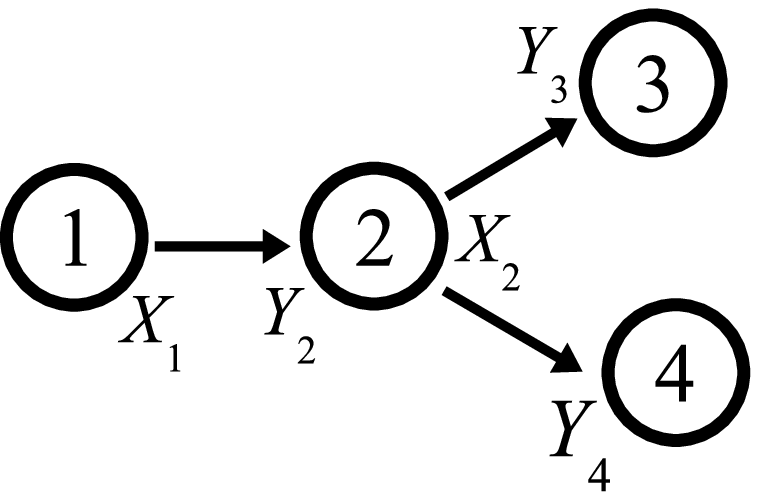}
\caption{One-hop myopic decode-forward for the four-node broadcast relay channel.} \label{fig:brc-myopic}
\end{minipage}
\end{figure}

\subsubsection{Omniscient Coding}
In omniscient decode-forward for the four-node broadcast relay channel, node 1 transmits to nodes 2, 3, and 4, while node 2 transmits to nodes 3 and 4. This is depicted in Fig.~\ref{fig:brc-omniscient}. Kramer \emph{et. at}~\cite{kramergastpar04b} gives achievable rates for the case where there are independent individual messages for nodes 3 and 4 as well as common messages for both receivers. In this paper, we consider the case where there is no private message. Under this condition, the following common rates~\cite[eq. (28)]{kramergastpar04b} are achievable by omniscient decode-forward.
\begin{equation}
R_0 \leq \min [ I(X_1;Y_2|X_2) , I(X_1,X_2;Y_3) , I(X_1,X_2;Y_4) ] = R_\text{omniscient}\label{eq:broadcast relay channel_omni}.
\end{equation}
Similar to the multiple-access relay channel, omniscient decode-forward is equivalent to two-hop decode-forward for the four-node broadcast relay channel.

\subsubsection{One-Hop Myopic Coding}
In one-hop myopic decode-forward for the four-node broadcast channel, node 1 transmits to only node 2, and node 2 transmits to nodes 3 and 4. This is depicted in Fig.~\ref{fig:brc-myopic}.  This is equivalent to a single point-to-point channel cascaded with a broadcast channel. The following rates are achievable by one-hop myopic decode-forward.
\begin{subequations}
\begin{align}
R_0 & \leq \min [ I(U_0;Y_3) , I(U_0;Y_4) ] \label{eq:brc-bc-1}\\
R_0 + R_3 & \leq I(U_0;Y_3) + I(U_3;Y_3|U_0) =  I(U_0,U_3;Y_3)\\
R_0 + R_4 & \leq I(U_0;Y_4) + I(U_4;Y_4|U_0) = I(U_0,U_4;Y_4)\\
R_0 + R_3 + R_4 & \leq \min [ I(U_0;Y_3) , I(U_0;Y_4) ] + I(U_3;Y_3|U_0) + I(U_4;Y_4|U_0) - I(U_3;U_4|U_0) \label{eq:brc-bc-2}\\
R_0 + R_3 + R_4 & \leq I(X_1;Y_2|X_2), \label{eq:brc-ptp}
\end{align}
\end{subequations}
for some $p(u_0,u_3,u_4,x_1,x_2) = p(x_1)p(u_0,u_3,u_4,x_2)$. The rates are be obtained by cascading a point-to-point channel (from node 1 to node 2) to a broadcast channel (from node 2 to nodes 3 and 4). Equation~\eqref{eq:brc-ptp} gives the rate constraints on the point-to-point channel; \eqref{eq:brc-bc-1}--\eqref{eq:brc-bc-2} gives the rate constraints on the broadcast channel with common information \cite[p. 391]{csiszatkorner81}. Here, ${U}_0$ carries information to be decoded by both nodes 3 and 4. ${U}_3$ and ${U}_4$ carry private information to nodes 3 and 4 respectively.
We set private messages to zero, that is $R_3=R_4=0$. We choose $U_0 = X_2, U_3=U_4=0$. Hence, the rate at which common messages can be sent to both receivers is
\begin{equation}
R_0 \leq \min [ I(X_1;Y_2|X_2), I(X_2;Y_3), I(X_2;Y_4) ] = R_\text{1-hop} \label{eq:broadcast relay channel_myopic}.
\end{equation}

We see that \eqref{eq:broadcast relay channel_myopic} differs from \eqref{eq:broadcast relay channel_omni} in the last two terms.  In the former, there is no cooperation between node 1 and node 2. In the latter, cooperation under the omniscient coding is reflected in the term $(X_1,X_2)$.

\subsection{Performance Comparison}
\subsubsection{Channel Setup}
We compare achievable rates of one-hop myopic decode-forward and omniscient decode-forward for the four-node Gaussian broadcast relay channel.  Nodes 2, 3, and 4 receive the following signal respectively.
\begin{subequations}
\begin{align}
Y_2 & = \sqrt{\kappa d_{12}^{-\eta}}X_1 + Z_2\\
Y_3 & = \sqrt{\kappa d_{13}^{-\eta}}X_1 + \sqrt{\kappa d_{23}^{-\eta}}X_2 + Z_3\\
Y_4 & = \sqrt{\kappa d_{14}^{-\eta}}X_1 + \sqrt{\kappa d_{24}^{-\eta}}X_2 + Z_4\\
\end{align}
\end{subequations}
where $E[X_1^2] = P_1$, $E[X_2^2] = P_2$, and $Z_2$, $Z_3$, and $Z_4$ are white Gaussian noise with variances $N_2$, $N_3$, and $N_4$ respectively.
In the analysis in this section, we use the following parameters: $d_{23} = d_{24} = d_{34} = 1$m, $d_{13} = d_{14}$, $N_2 = N_3 = N_4 = 1$W, $\kappa=1$, and $\eta=2$.

\subsubsection{One-Hop Myopic Coding}
In one-hop myopic decode-forward, the reception rate at node 2 is
\begin{subequations}
\begin{align}
R'_2 & = \frac{1}{2} \log 2\pi e [ \kappa d_{12}^{-\eta}P_1 + N_2] - \frac{1}{2} \log 2\pi e N_2\\
& = \frac{1}{2} \log \left[ 1 + \frac{P_1}{d_{12}^2} \right].
\end{align}
\end{subequations}
Due to symmetry, the reception rates at both node 3 and node 4 are
\begin{subequations}
\begin{align}
R'_3 = R'_4 & = \frac{1}{2} \log 2\pi e [ \kappa d_{23}^{-\eta}P_2 + \kappa d_{13}^{-\eta}P_1 + N_3 ]  - \frac{1}{2} \log 2\pi e [ \kappa d_{13}^{-\eta}P_1 + N_3 ] \\
& = \frac{1}{2} \log \left[ 1 + \frac{P_2/d_{23}^2}{1 + P_1/d_{13}^2} \right]\\
& = \frac{1}{2} \log \left[ 1 + \frac{P_2}{1 + \frac{P_1}{1/4 + ( \sqrt{3}/2 + d_{12} )^2} } \right].
\end{align}
\end{subequations}

Hence, achievable common rates are up to
\begin{subequations}
\begin{align}
R_0 &\leq \min \{ R_2', R_3', R_4'\}\\
& = \frac{1}{2} \log \left[ 1 + \min \left\{ \frac{P_1}{d_{12}^2} , \frac{P_2}{1 + \frac{P_1}{1/4 + ( \sqrt{3}/2 + d_{12} )^2} } \right\} \right]\\
& = R_\text{1-hop}.
\end{align}
\end{subequations}

\begin{figure}[t]
\begin{minipage}[b]{0.48\linewidth}
\centering
\includegraphics[width=\textwidth]{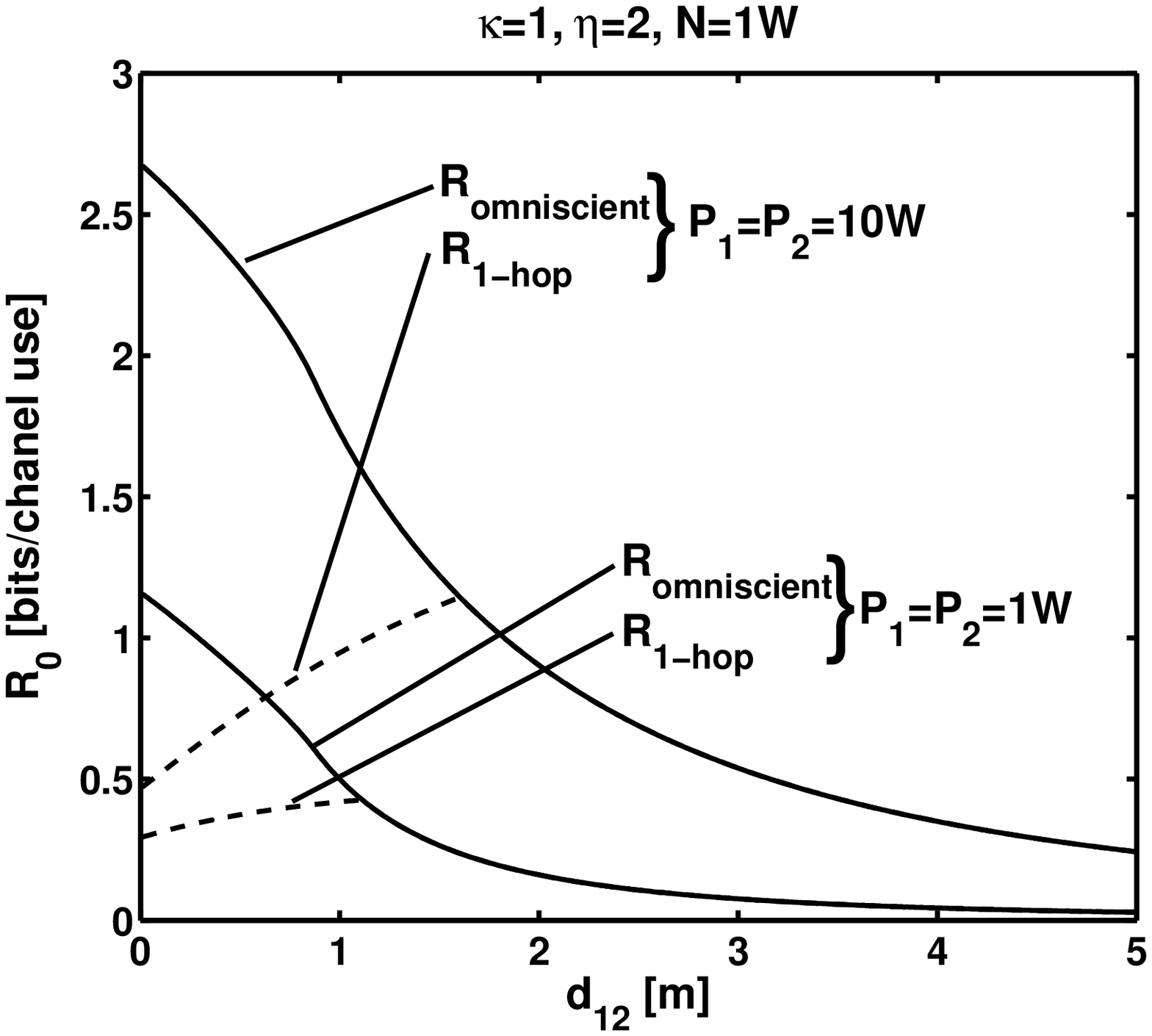}
\caption{$R_0$ vs. $d_{12}$ for one-hop myopic decode-forward and omniscient decode-forward for the four-node broadcast relay channel.} \label{fig:broadcast relay channel_rate_vs_d12}
\end{minipage}
\hspace{0.3cm}
\begin{minipage}[b]{0.48\linewidth}
\centering
\includegraphics[width=\textwidth]{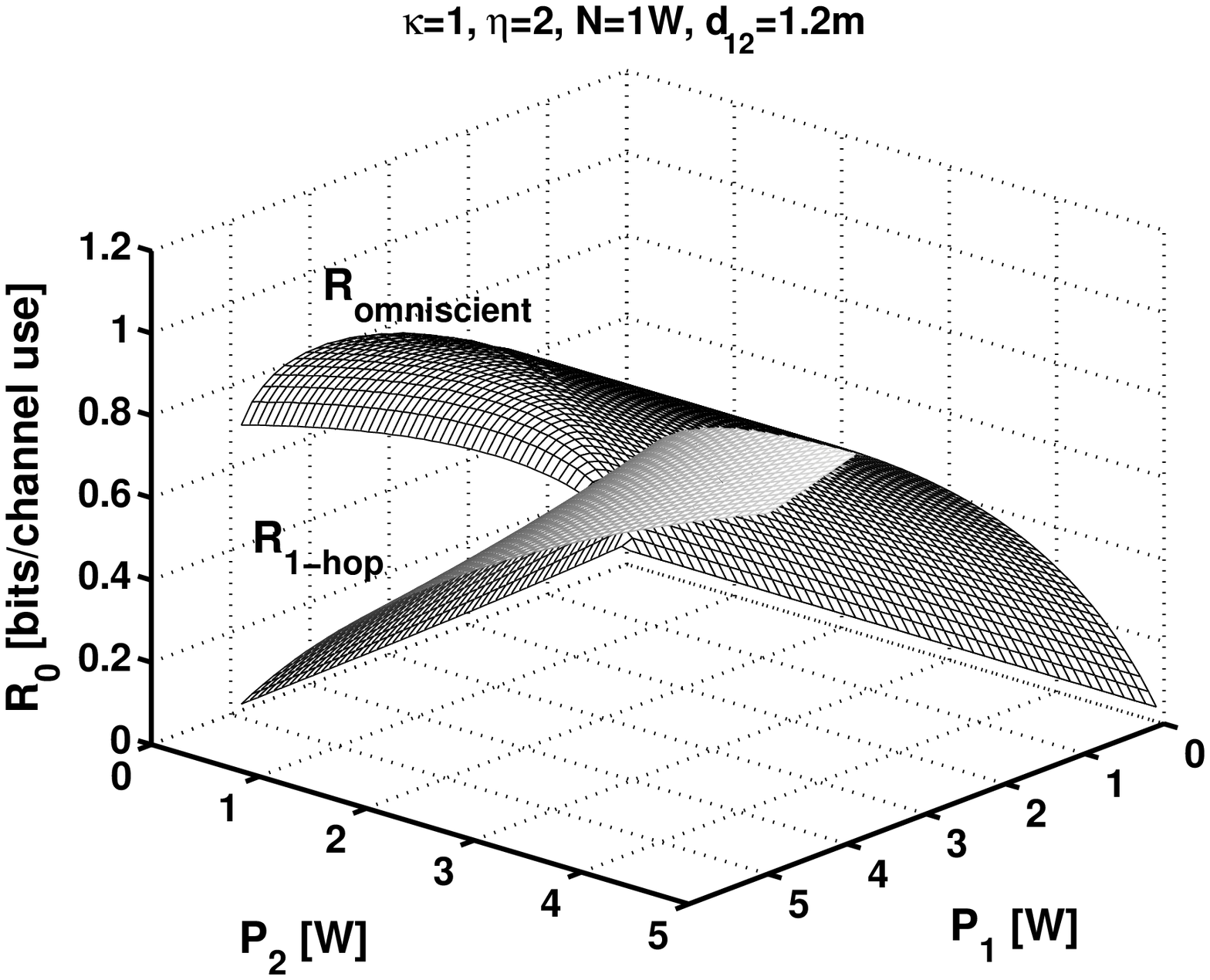}
\caption{Comparison of achievable sum rates of one-hop myopic decode-forward and omniscient decode-forward for the four-node broadcast relay channel.} \label{fig:broadcast relay channel_rate_vs_P1_vs_P2}
\end{minipage}
\end{figure}

\subsubsection{Omniscient coding}
In the case where only common messages are to be sent, the channel can be simplified to two identical relay channels due to symmetry.  Similar to the relay channel, nodes 1 and 2 transmit the following respectively.
\begin{subequations}
\begin{align}
X_1 & = \sqrt{P_1} ( \sqrt{\alpha}U_2 + \sqrt{1-\alpha}U_1 )\\
X_2 & = \sqrt{P_2}U_2
\end{align}
\end{subequations}
where $U_2$ and $U_1$ are independent zero-mean Gaussian random variables with unit variance.

The reception rate at node 2 is
\begin{subequations}
\begin{align}
R'_2 & = I(X_1;Y_2|X_2)\\
& = \frac{1}{2} \log 2\pi e [ \kappa d_{12}^{-\eta}\alpha P_1 + N_2 ] - \frac{1}{2} \log 2\pi e N_3\\
& = \frac{1}{2} \log \left[ 1 + \frac{ (1-\alpha) P_1 } { d_{12}^2}  \right]
\end{align}
\end{subequations}
and the reception rate at node 3 (and node 4 due to symmetry) is
\begin{subequations}
\begin{align}
R'_3 = R'_4 & = I(X_1,X_2; Y_3)\\
& = \frac{1}{2} \log 2\pi e \Bigg[ \kappa d_{13}^{-\eta}(1-\alpha) P_1 + \left( \sqrt{ \kappa d_{13}^{-\eta}\alpha P_1} + \sqrt{ \kappa d_{23}^{-\eta}P_2}\right)^2 + N_3 \Bigg] - \frac{1}{2} \log 2\pi e N_3\\
& = \frac{1}{2} \log \Bigg[ 1 + \frac{P_1}{1/4 + ( \sqrt{3}/2+d_{12} )^2} + P_2  + 2 \sqrt{ \frac{ \alpha P_1P_2 }{ 1/4 + (\sqrt{3}/2+d_{12})^2 } } \Bigg].
\end{align}
\end{subequations}

Hence, achievable common rates are up to
\begin{subequations}\label{eq:broadcast relay channel_omni_overall}
\begin{align}
R_0 &\leq \min \{ R_2', R_3', R_4'\}\\
& = \frac{1}{2} \log \Bigg( 1 + \min \Bigg\{ \frac{(1-\alpha) P_1}{d_{12}^2} , \frac{P_1}{d_{13}^2} + P_2 + 2 \sqrt{ \frac{ \alpha P_1P_2 }{ d_{13}^2 } } \Bigg\} \Bigg)\\
& = R_\text{omniscient},
\end{align}
\end{subequations}
for some $0 \leq \alpha \leq 1$, where $d_{13}^ 2 = 1/4 + ( \sqrt{3}/2+d_{12} )^2$.

In Fig.~\ref{fig:broadcast relay channel_rate_vs_d12}, the maximum achievable common rate is constrained by $R'_3$ (and $R'_4$) when $d_{12}$ is small, and by $R'_2$ when $d_{12}$ gets large.  From the rate expressions, we see that $R'_2$ of the myopic coding and the omniscient coding has the same expression (by setting $\alpha=0$ in the latter).  When the maximum achievable common rate is constrained by $R'_2$, the optimal $\alpha$ is 0, to make the first term in \eqref{eq:broadcast relay channel_omni_overall} largest possible.  When $d_{12}$ is large, $R_2'$ is the bottleneck, and achievable rates under both coding strategies are the same. This is because using either the myopic coding or the omniscient coding, node 2 only decodes from node 1.  Comparing the transmit power of 1W and 10W, when nodes transmit at lower power (or lower SNR) $R'_2$ constrains the overall rate for a larger range of $d_{12}$. So, achievable rates of one-hop myopic decode-forward are as high as that of omniscient decode-forward for larger range of $d_{12}$ in the low SNR regime.

Fig.~\ref{fig:broadcast relay channel_rate_vs_P1_vs_P2} depicts  achievable rates of one-hop myopic decode-forward and that of omniscient decode-forward for different $P_1$ and $P_2$. Achievable rates of the myopic coding are as high as that of the omniscient coding when $P_1$ is low and $P_2$ is high.  This is exactly the criteria for $R_0$ to be constrained by $R'_2$, or in other words, when the source-relay link is the bottleneck.

\section{Conclusion}\label{sec:myopic-conclusion}
We derived achievable rates of myopic decode-forward coding strategies for the multiple-relay channel, the multiple-access relay channel, and the broadcast relay channel. Myopic coding has practical advantages of being more robust to network topology changes, less processing, and fewer storage requirements at each node.

We showed that in the low SNR regime, achievable rates of two-hop myopic decode-forward are as large as that of omniscient decode-forward in a five-node multiple-relay channel, and close to that of the omniscient coding in a six-node channel. Comparing one-hop myopic decode-forward and two-hop myopic decode-forward, we see that adding a node into the nodes' view improves the achievable rate significantly. Hence, besides being more practical, a myopic coding strategy potentially (as only non-constructive coding is being considered) performs as good or close to the corresponding omniscient coding strategy. This means in a large network, we might do local coding design without compromising much on the achievable rate.

We also analyzed two myopic coding strategies in the multiple-access relay channel and the broadcast relay channel.  Using examples of four-node Gaussian channels, we showed that achievable rates of these myopic coding strategies are as good as that of their corresponding omniscient coding strategies when the source(s) transmit(s) at low power and the relay transmits at high power.

The analysis in this paper helps us to understand coding in multi-terminal networks better.  This work sheds light on the
practical design of efficient transmission protocols in wireless
networks, where robustness, computational power, and storage memory
are important design considerations, in addition to transmission rate.

\appendices

\section{An Example to Show that Myopic Coding is More Robust} \label{myopic-robustness-example}
To illustrate the robustness of myopic coding, we consider decode-forward in the seven-node Gaussian multiple-relay network in which node 4 fails. This means the signal contributed by node 4 will stop. We consider the following scenarios in myopic and omniscient coding:
\begin{enumerate}
\item Two-hop myopic decode-forward:
\begin{enumerate}
\item When the overall transmission rate is not affected: Node 2 decodes only from node 1, and cancels the interference only from itself (echo cancellation) and node 3. So, the failure of node 4 does not affect the decoding at node 2. Node 7 will also not be affected as it decodes only from nodes 5 and 6. In brief, the failure of node $t$ only affects nodes $t-1, t+1$, and $ t+2$ in two-hop myopic decode-forward.
\item When the overall transmission rate is affected: Suppose that upon node 4's failure, the overall transmission rate is lowered due to the change in the reception rate of node 5. Additional re-configuration at the source is required. Now, the source will have to transmit at a lower rate. One way of doing this is to use the existing code, but pad the lower rate messages with zeros. With zero-padding, the encoding and decoding at nodes 2  and 7 need not be changed as the supported rates at these nodes are not affected.
\end{enumerate}
\item Omniscient decode-forward: Nodes 2 and 3, who presume that node 4 is still transmitting and attempt to cancel its transmissions, will introduce more noise to their decoders. Nodes 5 to 7, who use node 4's signal contribution in the decoding, will experience a lower SNR. Hence the supported rates at these nodes will be lowered.
\end{enumerate}
Using omniscient coding, any topology change in the network (e.g., node failure or relocation) requires re-configuration of more nodes compared to using myopic coding.

\section{Proof of Theorem~\ref{thm:two-hop_myopic}}\label{append:two-hop}
In this appendix, we describe the encoding and decoding schemes, and prove achievable rates of two-hop myopic decode-forward for the multiple-relay channel. We consider $B+T-2$ transmission blocks, each of $n$ uses of the channel. A sequence of independent $B$ indices, $w_b
\in \{ 1, 2, \dotsc, 2^{nR} \}$, $b = 1, 2, \dotsc, B$ are sent
over $n(B+T-2)$ uses of the channel.  As $B \rightarrow \infty$, the
rate $RnB/n(B+T-2) \rightarrow R$ for any $n$.

\emph{Note:} We use $w$ and $z$ to represent the source message. The notation $w_j$ denotes the information which the source outputs at the $j$-th block. This means the source emits $w_1, w_2, \dotsc$ in blocks $1, 2, \dotsc$ respectively. The notation $z_t$ denotes the new information which node $t$ transmits. Since each node transmits codewords derived from the last two decoded messages, node 2 always transmits $(z_2,z_3)$. These different notations are used at different instances for better illustration.

\subsection{Codebook Generation}
In this section, we see how the codebook at each node is generated.

\begin{itemize}
\item First, fix the probability distribution
\begin{equation}
p(u_1,u_2,\dotsc,u_{T-1},x_1,x_2,\dotsc,x_{T-1}) = p(u_1)p(u_2)\dotsm p(u_{T-1}) p(x_1|u_1,u_2) p(x_2|u_2,u_3) \dotsm p(x_{T-1}|u_{T-1}) \nonumber
\end{equation}
for each $u_i \in \mathcal{U}_i$.
\item For each $t \in \{1,\dotsc,T-1\}$, generate $2^{nR}$ independent and identically distributed (i.i.d.) $n$-sequences in $\mathcal{U}_t^n$, each drawn according to $p(\mathbf{u}_{t}) = \prod_{i=1}^n p(u_{ti})$.  Index them as $\mathbf{u}_{t}(z_{t})$, $z_{t} \in \{1,\dotsc, 2^{nR}\}$.
\item Define $\mathbf{x}_{T-1}(z_{T-1}) = \mathbf{u}_{T-1}(z_{T-1})$.
\item For each $t \in \{1,\dotsc,T-2\}$, define a deterministic function that maps $(\mathbf{u}_t,\mathbf{u}_{t+1})$ to $\mathbf{x}_t$:
\begin{equation}
\mathbf{x}_t(z_t,z_{t+1}) =
f_t \big( \mathbf{u}_t(z_t),\mathbf{u}_{t+1}(z_{t+1}) \big).
\end{equation}
\item Repeat the above steps to generate a new independent codebook \cite{xiekumar03}.  These two codebooks are used in alternate block of transmission.  The reason for using two independent codebooks will be clear in the error probability analysis section.
\end{itemize}

We see that in each transmission block, node $t$, $t \in \{1,\dotsc,T-2\}$,
sends messages of two blocks: $z_t$ (new data) and $z_{t+1}$ (old
data). In the same block, node $t+1$ sends messages $z_{t+1}$ and
$z_{t+2}$. Note that a node cooperates with the node in the next hop
by repeating the transmission $z_{t+1}$.  We will see this clearer in the next
section.

\subsection{Encoding}
\begin{figure}[t]
\centering
\includegraphics[width=10cm]{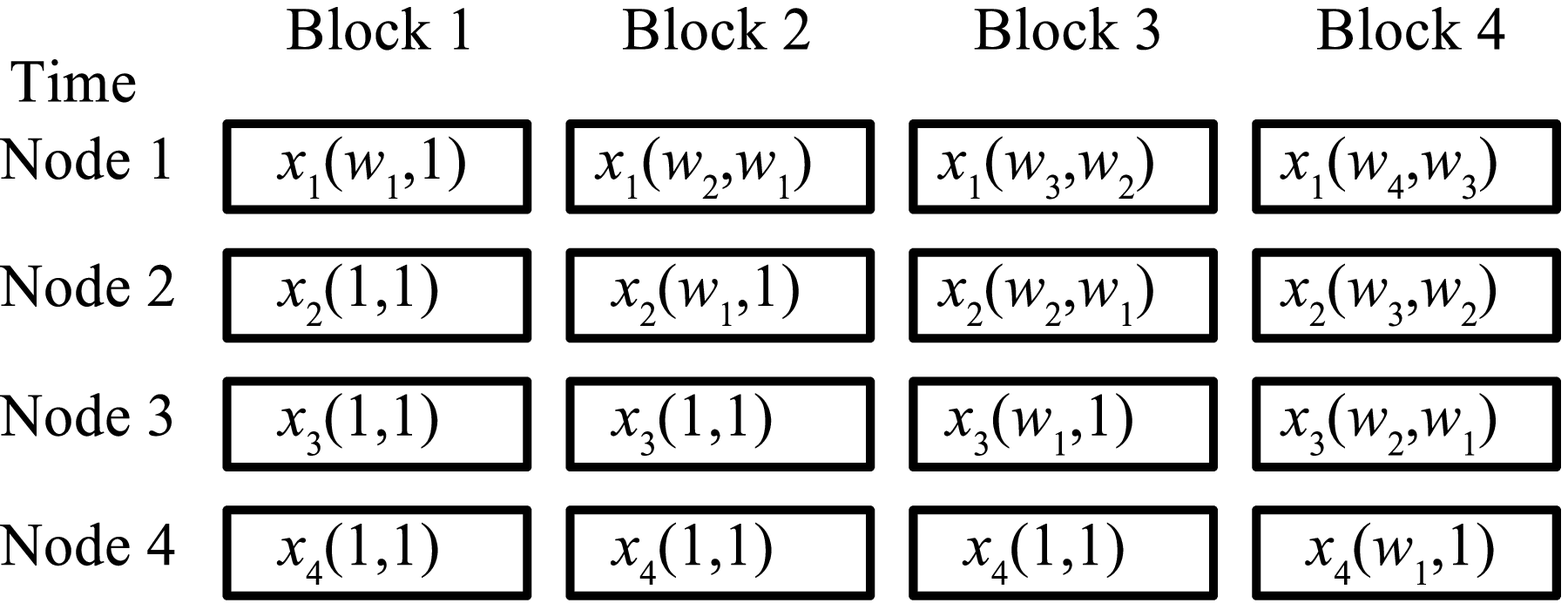}
\caption{The encoding scheme of two-hop myopic decode-forward for the multiple-relay channel.}
\label{fig:two_hop_encoding}
\end{figure}

Fig.~\ref{fig:two_hop_encoding} shows the encoding process for two-hop myopic decode-forward. The encoding steps are as follows:
\begin{itemize}
\item In the beginning of block 1, the source emits the first source letter $w_1$. Note that there is no new information after $B$ blocks.  We define $w_{B+1} = w_{B+2} = \dotsm = w_{B+T-2} = 1$.
\item In block 1, node 1 transmits $\mathbf{x}_1(w_1,w_0)$. Since the rest of the nodes have not received any information, they send dummy symbols $\mathbf{x}_i(w_{2-i},w_{1-i})$, $i \in \{2,\dotsc,T-1\}$.  We define $w_b=1$, for $b \leq 0$. In block 1, $z_1 = w_1, z_2 = w_0, \dotsc$
\item At the end of block 1, assume that node 2 correctly decodes the first signal $w_1$.
\item In block 2, node 2 transmits $\mathbf{x}_2(w_1,w_0)$.  Node 1 transmits $\mathbf{x}_1(w_2,w_1)$.  It helps node 2 to re-transmit $w_1$ and sends $w_2$ (new information) at the same time. In block 2, $z_1 = w_2, z_2 = w_1, z_3= w_0, \dotsc$
\item Generalizing, in block $b \in \{1,\dotsc,B+T-2\}$, node $t$, $t \in \{1,\dotsc,T-1\}$, has data $(w_1, w_2, \dotsc, w_{b-t+1})$. Under two-hop myopic decode-forward, it sends $\mathbf{x}_t(w_{b-t+1},w_{b-t})$. 
\item We see that a node sends messages that it has decoded in the past two blocks. This adheres to the constraints of two-hop myopic decode-forward.
\end{itemize}

\subsection{Decoding}
\begin{figure}[t]
\centering
\includegraphics[width=12cm]{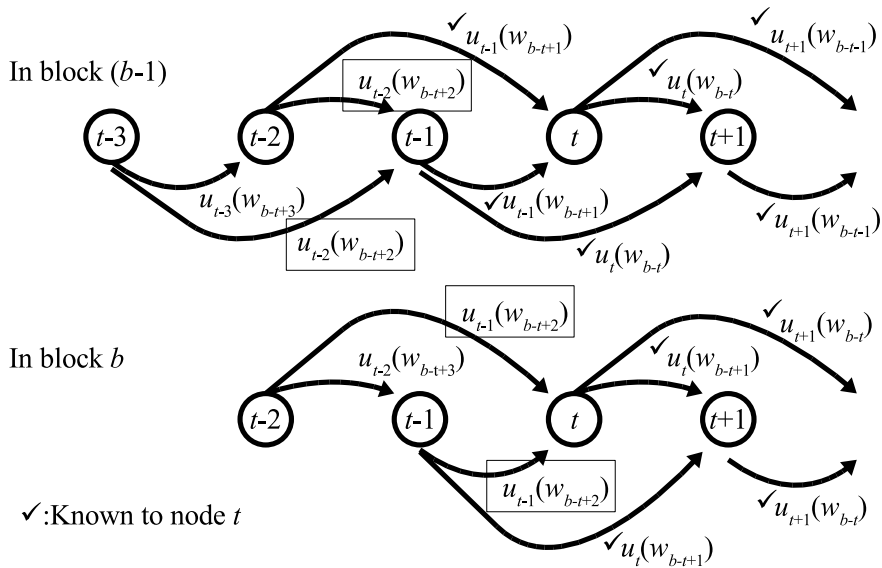}
\caption{Decoding at node $t$ of message $w_{b-t+2}$.}\label{fig:decoding_ex}
\end{figure}

\begin{itemize}
\item Under the two-hop myopic decode-forward constraints, a node can store a decoded message no longer than two blocks and can use two blocks of received signal to decode one message.
\item Node 2's decoding is slightly different from the other nodes as it has only one upstream node. So it decodes every message using one block of received signal. We illustrate the decoding of message $w_4$ at node 2. At the end of block 4, assuming that node 2 has already decoded messages $(w_1,w_2,w_3)$ correctly. However, due to the myopic coding constraint, it only has $w_2$ and $w_3$ in its memory. This is because $w_1$ was decoded at the end of block 1 and would have to be discarded at the end of block 3. So, it finds the a unique $\mathbf{u}_1(w_4)$ which is jointly typical with $\mathbf{u}_3(w_2), \mathbf{u}_2(w_3),$ and $\mathbf{y}_{2,4}$ (the received signal at node 2 in block 4). We write $\mathbf{y}_{2,4}$ instead of $\mathbf{y}_{24}$ to avoid the confusion with the received signal of node 24. An error is declared is there if no such $w_4$ or more than one unique $w_4$.
\item Nodes 3 to $T$ decode a message using two blocks of received signal. Consider node 3. At the end of block 4, assuming that node 3 has already decoded $w_1$ (decoded at the end of block 2) and $w_2$ (decoded at the end of block 3) correctly. Assume that it now correctly decodes $w_3$ using signals from blocks 3 and 4. At the end of block 4, it finds a set of $\mathbf{u}_1(w_4)$ which is jointly typical with $\mathbf{u}_4(w_1), \mathbf{u}_3(w_2), \mathbf{u}_2(w_3),$ and $\mathbf{y}_{3,4}$. We call this set $\mathcal{L}_1(w_4)$. Since it can only keeps messages decoded over two blocks, it keeps $w_2$ and $w_3$ and discard $w_1$. At the end of block 5, node 3 finds a set of $\mathbf{u}_2(w_4)$ that is jointly typical with $\mathbf{u}_4(w_2), \mathbf{u}_3(w_3),$ and $\mathbf{y}_{3,5}$. We call this set $\mathcal{L}_2(w_4)$. It finds a unique $w_4$ that belong to both sets, that is $\hat{w}_4 \in \mathcal{L}_1(w_4) \cap \mathcal{L}_2(w_4)$. Here $\cap$ denotes intersection of sets. An error is declared when the intersection contains more than one index or the sets do not intersect.
\item We now generalize the decoding process. Refer to Fig.~\ref{fig:decoding_ex}, at the end of block $b-1$, assuming that node $t$ has correctly decoded $(w_1, \dotsc, w_{b-t})$. Under the myopic coding constraint, it has in its memory $w_{b-t-1}$ and $w_{b-t}$. It decodes $w_{b-t+1}$. It then finds a set of $\mathbf{u}_{t-2}(w_{b-t+2})$ that is jointly typical with $(\mathbf{u}_{t-1}(w_{b-t+1}),$ $\mathbf{u}_t(w_{b-t}), \mathbf{u}_{t+1}(w_{b-t-1}), \mathbf{y}_{t(b-1)})$. Label this set $\mathcal{L}_1(w_{b-t+2})$. It discards $w_{b-t-1}$ from its memory.  At the end of block $b$, it finds the set of $\mathbf{u}_{t-1}(w_{b-t+2})$ that is jointly typical with $(\mathbf{u}_t(w_{b-t+1}), \mathbf{u}_{t+1}(w_{b-t}), \mathbf{y}_{tb})$. Label this set $\mathcal{L}_2(w_{b-t+2})$. It declare $\hat{w}_{b-t+2}$ if there is one and only one index in $\mathcal{L}_1(w_{b-t+2}) \cap \mathcal{L}_2(w_{b-t+2})$.
\end{itemize}

\subsection{Achievable Rates and Probability of Error Analysis}
In the previous section, we said that node $t$ decodes message $w_{b-t+2}$ in block $b$.  We denote the event that no decoding error is made at all nodes in the first $b$ block, $1 \leq b \leq B+T-2$, by
\begin{equation}
\mathcal{C}(b) \triangleq \{ \hat{w}_{t(k-t+2)} = w_{k-t+2}:
\forall t \in [2,T] \text{ and } k \in [1,b] \}
\end{equation}
where $\hat{w}_{t(b)}$ is node $t$'s estimate of the message $w_b$.
This means in the first $b$ blocks, node 2 will have correctly
decoded $(w_1, w_2, \dotsc, w_b)$, node $3$ will have correctly
decoded $(w_0, w_1, \dotsc, w_{b-1})$, and so on.  We set $w_k=1$
for $k \leq 0$.  They are the dummy signals sent by the nodes.

We denote the probability that there is no decoding error up to
block $b$ as
\begin{equation}
P_c(b) \triangleq \Pr \{ \mathcal{C}(b) \}
\end{equation}
and $P_c(0) \triangleq 1$.  The probability that one or more error occurs during block $b \in [1, B+T-2]$ at some node $t \in [2,T]$, given that there is no error in decoding at all nodes in all blocks up to $b-1$, is
\begin{subequations}
\begin{align}
P_e(b)\nonumber
& \triangleq \Pr \Big\{ \hat{w}_{t(b-t+2)} \neq w_{b-t+2}: 
\text{ for some } t \in \{2,\dotsc,T\} \Big| \mathcal{C}(b-1) \Big\}\\
& \leq \sum_{t=2}^T \Pr \left\{ \hat{w}_{t(b-t+2)} \neq w_{b-t+2} |
\mathcal{C}{(b-1)} \right\}\\
& \triangleq \sum_{t=2}^T P_{et}(b)
\end{align}
\end{subequations}
where $P_{et}(b) \triangleq \Pr \left\{ \hat{w}_{t(b-t+2)} \neq
w_{b-t+2} | \mathcal{C}{(b-1)} \right\}$, which is the probability
that node $t$ wrongly decodes the latest letter $w_{b-t+2}$ in block
$b$, given that it has correctly decoded the past letters.

Now, we need to compute the error probability $P_{et}(b)$. As
mentioned in the decoding section, the decoding of a message spans
over two blocks.  For example, let us look at the decoding of message
$w_{b-t+2}$ at node $t$, as depicted in
Fig.~\ref{fig:decoding_ex}. The message to be decoded is boxed and
the messages that node $t$ has correctly decoded are marked with
$\checkmark$.  In block $b-1$, node $t$ find a set of $w_{b-t+2}$
for which
\begin{equation}
\Big( \mathbf{u}_{t-2}(w_{b-t+2}), \mathbf{u}_{t-1}(w_{b-t+1}),
\mathbf{u}_t(w_{b-t}), \mathbf{u}_{t+1}(w_{b-t-1}),
\mathbf{y}_{t(b-1)}
\Big) \in \mathcal{A}_\epsilon^n (U_{t-2}, U_{t-1},U_t,U_{t+1},Y_t)
\triangleq \mathcal{A}_1.
\end{equation}
In block $b$, node $t$ finds a set of $w_{b-t+2}$ for which
\begin{equation}
\left( \mathbf{u}_{t-1}(w_{b-t+2}), \mathbf{u}_{t}(w_{b-t+1}),
\mathbf{u}_{t+1}(w_{b-t}), \mathbf{y}_{tb} \right) \in \mathcal{A}_\epsilon^n (U_{t-1},U_t,U_{t+1},Y_{t}) \triangleq
\mathcal{A}_2.
\end{equation}
Node $t$ then finds the intersection of the two sets to determine the
value of $w_{b-t+2}$.

Assuming that node $t$ has correctly decoded
$w_{b-t-1}$, $w_{b-t}$, and $w_{b-t+1}$, we define the following error
events:
\begin{subequations}
\begin{align}
\mathcal{E}_1 & \triangleq \Big( \mathbf{u}_{t-2}(w_{b-t+2}),
\mathbf{u}_{t-1}(w_{b-t+1}), \mathbf{u}_t(w_{b-t}),
\mathbf{u}_{t+1}(w_{b-t-1}),\mathbf{y}_{t(b-1)} \Big) \notin
\mathcal{A}_1\\
\mathcal{E}_2 & \triangleq \Big( \mathbf{u}_{t-2}(v),
\mathbf{u}_{t-1}(w_{b-t+1}), \mathbf{u}_t(w_{b-t}),
\mathbf{u}_{t+1}(w_{b-t-1}), \mathbf{y}_{t(b-1)} \Big) \in
\mathcal{A}_1\label{eq:A1_b}\\
\mathcal{E}_3 & \triangleq \Big( \mathbf{u}_{t-1}(w_{b-t+2}),
\mathbf{u}_{t}(w_{b-t+1}), \mathbf{u}_{t+1}(w_{b-t}), \mathbf{y}_{tb}
\Big) \notin \mathcal{A}_2\\
\mathcal{E}_4 & \triangleq \Big( \mathbf{u}_{t-1}(v),
\mathbf{u}_{t}(w_{b-t+1}), \mathbf{u}_{t+1}(w_{b-t}), \mathbf{y}_{tb}
\Big) \in \mathcal{A}_2\label{eq:A2_b}
\end{align}
\end{subequations}
for some $v \in \left\{ v \in [1, \dotsc, 2^{nR}]: v \neq w_{b-t+2}
\right\}$, and
\begin{equation}
\mathcal{E}_5 \triangleq \mathcal{E}_2 \cap \mathcal{E}_4
\label{eq:A1A2}.
\end{equation}
$\mathcal{E}_5$ is the event where $v \neq w_{b-t+2}$ is found in the intersection of the
decoding sets and is, therefore, wrongly decoded as the transmitted message.  An
error occurs during the decoding in block $b$ at node $t$ if events $\mathcal{E}_1$,
$\mathcal{E}_3$, or $\mathcal{E}_5$ occurs.  Now, we can rewrite
\begin{equation}
P_{et}(b) = \Pr \{ \mathcal{E}_1 \cup \mathcal{E}_3 \cup
\mathcal{E}_5 \} \leq \Pr\{\mathcal{E}_1\} + \Pr\{\mathcal{E}_3\} +
\Pr\{\mathcal{E}_5\}.
\end{equation}
The last equation is due to the union bound of events.

From the definition of jointly typical sequences
(Definition~\ref{def:AEP}), we know that
\begin{subequations}
\begin{align}
\Pr \{ \mathcal{E}_1 \} & \leq \epsilon\\
\Pr \{ \mathcal{E}_3 \} & \leq \epsilon,
\end{align}
\end{subequations}
for sufficiently large $n$.

Using Lemma~\ref{lem:AEP}, we derive the probability of a particular
$v \neq w_{b-t+2}$ that satisfies \eqref{eq:A1_b}:
\begin{subequations}
\begin{align}
& \Pr \bigg\{ ( \mathbf{u}_{t-2}(v), \mathbf{u}_{t-1}(w_{b-t+1}),
\mathbf{u}_t(w_{b-t}), \mathbf{u}_{t+1}(w_{b-t-1}),\mathbf{y}_{t(b-1)} )
\in \mathcal{A}_1 \bigg\}\nonumber\\
& =
\sum_{(\mathbf{u}_{t-2},\mathbf{u}_{t-1},\mathbf{u}_{t},\mathbf{u}_{t+1},\mathbf{y}_t)\in\mathcal{A}_1}
p(\mathbf{u}_{t-2})p(\mathbf{u}_{t-1},\mathbf{u}_{t},\mathbf{u}_{t+1},\mathbf{y}_t)\\
& \leq |\mathcal{A}_1| 2^{-n(H(U_{t-2})-\epsilon)}
2^{-n(H(U_{t-1},U_{t},U_{t+1},Y_t)-\epsilon)}\\
& \leq 2^{n(H(U_{t-2},U_{t-1},U_{t},U_{t+1},Y_t)+\epsilon)}
2^{-n(H(U_{t-2})-\epsilon)}  2^{-n(H(U_{t-1},U_{t},U_{t+1},Y_t)-\epsilon)}\\
& = 2^{-n( H(U_{t-2}) - H(U_{t-2}|Y_t,U_{t-1},U_{t},U_{t+1})
-3\epsilon)}\\
& \leq 2^{-n( I(U_{t-2};Y_t|U_{t-1},U_t,U_{t+1}) - 3\epsilon)}.
\end{align}
\end{subequations}
The last equation is because $H(U_{t-2}) \geq H(U_{t-2} | U_{t-1},U_t,U_{t+1})$.

By a similar method, we can calculate the probability of a particular
$v \in \{v \in \{1,\dotsc,2^{nR}\}: v \neq w_{b-t+2} \}$ satisfies
\eqref{eq:A2_b}:
\begin{equation}
 \Pr \left\{ ( \mathbf{u}_{t-1}(v_2), \mathbf{u}_{t}(w_{b-t+1}),
\mathbf{u}_{t+1}(w_{b-t}), \mathbf{y}_{tb} ) \in \mathcal{A}_2
\right\} \leq 2^{-n( I(U_{t-1};Y_t|U_t,U_{t+1}) - 3\epsilon)}.
\end{equation}

Combining these two probabilities, we find the probability that node
$t$ wrongly decodes $w_{b-t+2}$ to any $v \in \{ v \in \{1,\dotsc,2^{nR}]: v
\neq w_{b-t+2} \}$ to be
\begin{subequations}
\begin{align}
&\Pr \{ \mathcal{E}_5\}\nonumber\\
 & = \sum_{\substack{v \in \{1,\dotsc,2^{nR}\}\\v \neq
w_{b-t+2}}} \Pr\{ v \text{ satisfies \eqref{eq:A1A2}} \}\\
& = \sum_{\substack{v \in \{1,\dotsc,2^{nR}\}\\v \neq w_{b-t+2}}} \Pr \{ v
\text{ satisfies \eqref{eq:A1_b}} \} \Pr \{ v \text{
satisfies \eqref{eq:A2_b}} \}\label{eq:indep_E5}\\
& \leq \left( 2^{nR} -1 \right) \times 2^{-n(
I(U_{t-2};Y_t|U_{t-1},U_t,U_{t+1}) - 3\epsilon)}
2^{-n( I(U_{t-1};Y_t|U_t,U_{t+1}) - 3\epsilon)}\\
& < 2^{-n(I(U_{t-2},U_{t-1};Y_t|U_t,U_{t+1})-6\epsilon - R)}\\
& \leq \epsilon.
\end{align}
\end{subequations}
Here, \eqref{eq:indep_E5} is due to the use of independent
codebooks for each alternating block.  The last equation is made
possible for sufficiently large $n$ and if
\begin{equation}\label{eq:two_hop_rate}
R < I(U_{t-2},U_{t-1};Y_t|U_t,U_{t+1})-6\epsilon.
\end{equation}
With this rate constraint and large $n$, we see that the probability
of error is
\begin{subequations}
\begin{align}
P_e(b) & = \sum_{t=2}^T P_{et}(b)\\
& \leq \sum_{t=2}^T \left[ \Pr\{\mathcal{E}_1\} +
\Pr\{\mathcal{E}_3\} + \Pr\{\mathcal{E}_5\} \right]\\
& \leq (T-1)3\epsilon,
\end{align}
\end{subequations}
which can be made arbitrarily small.  Hence, the rate in
\eqref{eq:two_hop_rate} is achievable.

Equation~\eqref{eq:two_hop_rate}  is
only the rate constraint at one node.  In two-hop myopic decode-forward,
each message must be fully decoded at each node, hence the overall
rate is constrained by
\begin{equation}
R \leq \min_{t \in \{2,\dotsc,T\}} R_t,\\
\end{equation}
where
\begin{equation}\label{eq:recep_rate_2hop}
R_t = I(U_{t-2},U_{t-1};Y_t|U_t,U_{t+1})
\end{equation}
and $U_0 = U_T = U_{T+1} = 0$. Since the message can flow through the relays in any order. Hence we arrive at Theorem~\ref{thm:two-hop_myopic}.

\section{Proof of Theorem~\ref{thm:k_hop}}\label{append:k_hop}
Now, we prove Theorem~\ref{thm:k_hop}. We start by describing the codebook generation.  We send $B$ blocks of information over $B+T-2$ blocks of channel use.

\subsection{Codebook Generation}
The codebook generation for $k$-hop myopic decode-forward for the multiple-relay channel is as follows.
\begin{itemize}
\item Fix the probability distribution function
\begin{subequations}
\begin{align}
& p(u_1, u_2, \dotsc, u_{T-1}, x_1, x_2, \dotsc, x_{T-1})\nonumber\\
& = p(u_1)p(u_2)\dotsm p(u_{T-1})p(x_{T-1}|u_{T-1}) \nonumber \\
& \quad \times p(x_{T-2}|u_{T-2},u_{T-1}) \dotsm p(x_{T-k}|u_{T-k},u_{T-k+1}\dotsc,u_{T-1})\nonumber\\
& \quad \times p(x_{T-k-1}|u_{T-k-1},u_{T-k} \dotsc, u_{T-2}) \dotsm p(x_1|u_1,u_2,\dotsc,u_k).
\end{align}
\end{subequations}
\item For each $t \in \{1,\dotsc,T-1\}$, generate $2^{nR}$ independent and identically distributed (i.i.d.) $n$-sequences in $\mathcal{U}_t^n$, each drawn according to $p(\mathbf{u}_{t}) = \prod_{i=1}^n p(u_{ti})$.  Index them as $\mathbf{u}_{t}(z_{t})$, $z_{t} \in \{1,\dotsc,2^{nR}\}$.
\item Define $\mathbf{x}_{T-1}(z_{T-1}) = \mathbf{u}_{T-1}(z_{T-1})$.
\item For each $t \in [T-k, T-2]$, define a deterministic function that maps $(\mathbf{u}_t,\mathbf{u}_{t+1},\dotsc,\mathbf{u}_{T-1})$ to $\mathbf{x}_t$:
\begin{equation}
\mathbf{x}_t(z_t,z_{t+1},\dotsc,z_{T-1}) = f_t \big( \mathbf{u}_t(z_t),\mathbf{u}_{t+1}(z_{t+1}),\dotsc,\mathbf{u}_{T-1}(z_{T-1}) \big).
\end{equation}
\item For each $t \in [1, T-k-1]$, define a deterministic function that maps $(\mathbf{u}_t,\mathbf{u}_{t+1},\dotsc,\mathbf{u}_{t+k-1})$ to $\mathbf{x}_t$:
\begin{equation}
\mathbf{x}_t(z_t,z_{t+1},\dotsc,z_{t+k-1}) = f_t \big( \mathbf{u}_t(z_t),\mathbf{u}_{t+1}(z_{t+1}),\dotsc,\mathbf{u}_{t+k-1}(z_{t+k-1}) \big).
\end{equation}
\item Repeat the above steps to generate $k-1$ new independent codebooks.  These $k$ codebooks are used in cycle and reused after $k$ blocks of $n$ transmissions.
\end{itemize}

For the sake of illustration, we denote the code of node $t$, $t \in \{1,\dotsc,T-1\}$ by $\mathbf{x}_t(z_t,z_{t+1},\dotsc,z_{t+k-1})$ where $z_j = 1$ for $j \geq T$.  These are dummy symbols that do not affect the encoding process.

\begin{figure}[t]
\centering
\includegraphics[width=12cm]{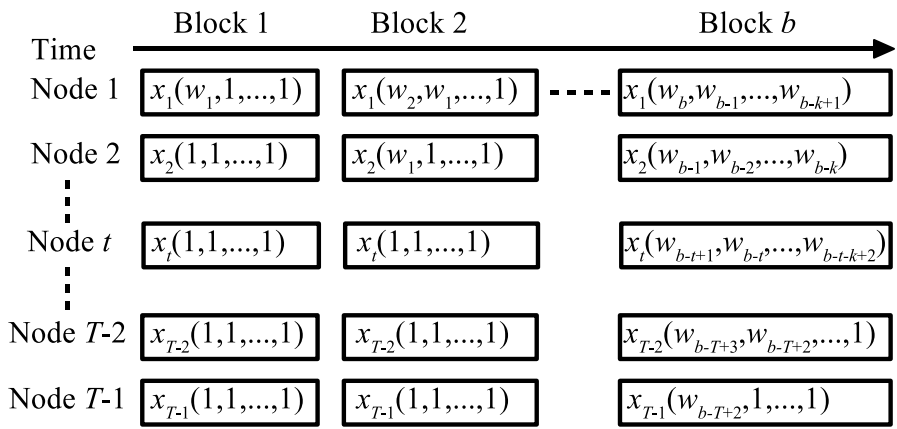}
\caption{The encoding scheme for $k$-hop myopic decode-forward for the multiple-relay channel.} \label{fig:k_hop_cod}
\end{figure}

\subsection{Encoding}
We now describe the encoding process for $k$-hop myopic decode-forward. It is depicted in Fig.~\ref{fig:k_hop_cod}.
\begin{itemize}
\item In the beginning of block 1, the source emits the first source letter $w_1$. Note that there is no new information in blocks $b$ for $B+1 \leq b \leq B+T-2$.  We assume that $w_{B+1} = w_{B+2} = \dotsm = w_{B+T-2} = 1$.
\item In block 1, node 1 transmits $\mathbf{x}_1(w_1,w_0,\dotsc,w_{2-k})$. Since the rest of the nodes have not received any information, they send dummy symbols $\mathbf{x}_i(w_{2-i},w_{1-i},\dotsc,w_{3-k-i})$, $i \in \{2,\dotsc,T-1\}$.  We define $w_b=1$, for $b \leq 0$.
\item At the end of block $b-1$, $b \geq 2$, we assume that node $t$ has correctly decoded messages up to $w_{b-t+1}$.  Under the $k$-hop myopic constraints, a node can encode with at most $k$ previously decoded messages in each block of transmission.  So, in block $b$, node $t$ encode $\min\{k,T-t\}$ previously decoded messages, i.e., it sends $\mathbf{x}_t(w_{b-t+1},w_{b-t},\dotsc,w_{b-t-k+2})$.  We note that there are only $T-t$ nodes in front of node $t$.  For the case of $T-t < k$, node $t$ sends $\mathbf{x}_t(w_{b-t+1},w_{b-t},\dotsc,w_{b-T+2},1, \dotsc, 1)$.  This means, it sets $w_i=1$ for $i \geq b-T+1$, which is equivalent to sending dummy symbols. This is because at the end of block $b-1$, node $T$ will have already correctly decoded signals up to $w_{b-T+1}$.  As this is the last node in the network, all other nodes will have had decoded those signals. Hence no node needs to transmit $w_i=1$ for $i \geq b-T+1$ again. The dummy symbols are included so that the same transmit notation can be used for all the nodes.
\end{itemize}

\begin{figure*}[t]
\centering
\includegraphics[width=16cm]{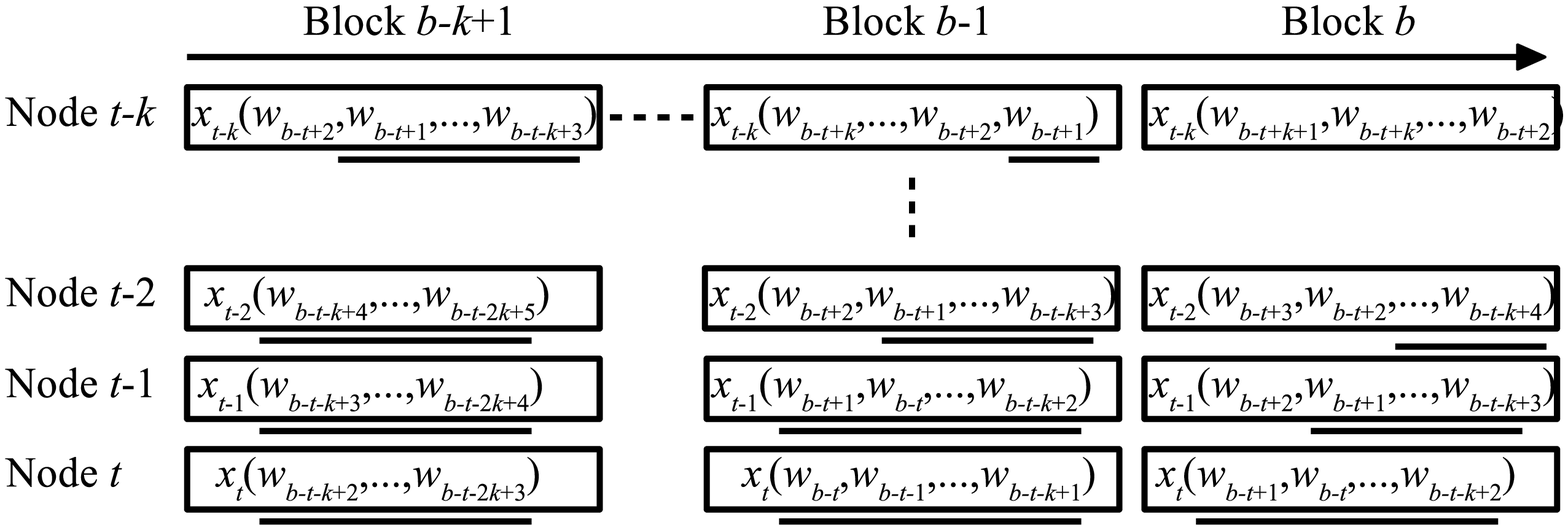}
\caption{The decoding scheme for $k$-hop myopic decode-forward for the multiple-relay channel.  Underlined symbols are those that has been decoded by node $t$ prior to block $b$.} \label{fig:k_hop_dec}
\end{figure*}

\subsection{Decoding and Achievable Rates of $k$-Hop Myopic Decode-Forward}
We look at how node $t$, for $t \geq k+1$, decodes $w_{b-t+2}$ at the end of block $b$. Fig.~\ref{fig:k_hop_dec} shows what the nodes transmit.
\begin{itemize}
\item During block $b$, there are $k$ nodes that encode $w_{b-t+2}$ in their transmissions. These are nodes $\{t-k, \dotsc, t-1\}$. Nodes $\{1, \dotsc, t-k-1\}$ do not encode $w_{b-t+2}$ in their transmissions in block $b$ as they have to discard the message due to the buffering constraint of the $k$-hop myopic coding.
\item At the end of block $b$, node $t$ finds $\mathcal{L}_1(\hat{w}_{b-t+2})$ in which
\begin{equation}\label{eq:contri_1}
\Big( \mathbf{u}_{t-1}(\hat{w}_{b-t+2}), \mathbf{u}_{t}(w_{b-t+1}), \dotsc, \mathbf{u}_{t+k-1}(w_{b-t-k+2}), \mathbf{y}_{tb} \Big) \in \mathcal{A}_\epsilon^n.
\end{equation}
Here, we note that node $t$ can store $k$ old messages.  Hence, during the decoding at the end of block $b$, it knows $\left(\mathbf{u}_t(w_{b-t+1}),\dotsc,\mathbf{u}_{t+k-1}(w_{b-t-k+2})\right)$. The rate contribution from \eqref{eq:contri_1} is
\begin{equation}
R_t^{(1)} = I(U_{t-1};Y_t|U_t,\dotsc,U_{t+k-1}).
\end{equation}
\item Moving back one block, at the end block $b-1$, 	node $t$ has messages $\left( \mathbf{u}_t(w_{b-t}), \dotsc, \mathbf{u}_{t+k-1}(w_{b-t-k+1}) \right)$ in its storage. After decoding $\mathbf{u}_{t-1}(w_{b-t+1})$, it then forms the set $\mathcal{L}_2(\hat{w}_{b-t+2})$ which
\begin{equation}
\Big( \mathbf{u}_{t-2}(\hat{w}_{b-t+2}), \mathbf{u}_{t-1}(w_{b-t+1}), \dotsc, \mathbf{u}_{t+k-1}(w_{b-t-k+1}), \mathbf{y}_{t(b-1)} \Big) \in \mathcal{A}_\epsilon^n.
\end{equation}
The rate contribution from this is
\begin{equation}
R_t^{(2)} = I(U_{t-2};Y_t|U_{t-1},\dotsc, U_{t+k-1}).
\end{equation}
\item Repeating this for blocks $(b-i+1)$, $3 \leq i \leq k$, node $t$ find the set $\mathcal{L}_i(\hat{w}_{b-t+2})$, and the rate contribution is
\begin{equation}
R_t^{(i)} = I(U_{t-i};Y_t|U_{t-i+1},\dotsc,U_{t+k-1}).
\end{equation}
The proof is similar to that for two-hop myopic decode-forward and will be omitted here.
\item Finally, node $t$ finds $\hat{w}_{b-t+2} \in \bigcap_{i=1}^{k} \mathcal{L}_i(\hat{w}_{b-t+2})$, where $\bigcap$ denotes the intersection of sets.  A unique $\hat{w}_{b-t+2}$ can be found if the reception rate at node $t$ is not more than
\begin{equation}
R_t = \sum_{i=1}^k R_t^{(i)} = I(U_{t-k},\dotsc,U_{t-1};Y_t|U_t,\dotsc ,U_{t+k-1}).
\end{equation}
\item Since all data must pass through every node, the overall rate is constrained by the node which has the lowest reception rate, that is
\begin{equation}
R \leq \min_{t \in \{ 2, \dotsc, T \}} R_t.
\end{equation}
\end{itemize}

With this, we have Theorem~\ref{thm:k_hop}.

\bibliography{bib}

\end{document}